%% file: main.tex
\documentclass[preprint,12pt]{elsarticle}

\usepackage[section]{placeins}
\usepackage{amssymb}

\usepackage{mathrsfs}
\usepackage{graphicx}
\usepackage{amsmath}
\usepackage{amsfonts}
\usepackage{amsthm}
\usepackage{textcomp}
\usepackage{color}
\usepackage{xcolor}
\usepackage[normalem]{ulem}
\usepackage{placeins}
\usepackage{bm}
\usepackage{wasysym}
\usepackage{geometry}
\usepackage{stmaryrd}
\usepackage{float}
\usepackage{subfigure}
\usepackage{pdfsync}
\usepackage[colorlinks,
            linkcolor=blue,
            anchorcolor=blue,
            citecolor=blue
            ]{hyperref}
\usepackage[ruled]{algorithm2e}
\usepackage{dotlessi}
\usepackage{caption}
\newtheorem{lemma}{Lemma}[section]

\definecolor{yscol}{HTML}{6622AA}

\newcounter{bla}

\journal{}

\begin{document}

\begin{frontmatter}



\title{MeshAC: A 3D Mesh Generation and Adaptation Package for Multiscale Coupling Methods\corref{equal}}
\cortext[equal]{\textbf{Funding}: YW is supported by Development Postdoctoral Scholarship for Outstanding Doctoral Graduates from Shanghai Jiao Tong University. LZ is supported by National Natural Science Foundation of China (NSFC 12271360, 11871339, 11861131004).}


\author[zjuaddress]{Kejie Fu}

\author[mingjieaddress]{Mingjie Liao}

\author[yangshuaiaddress]{Yangshuai Wang}
\author[zjuaddress]{Jianjun Chen}

\author[sjtuaddress]{Lei Zhang}

\address[zjuaddress]{Center for Engineering and Scientific Computation and School of Aeronautics and Astronautics, Zhejiang University, Hangzhou 310027, China.}

\address[mingjieaddress]{AsiaInfo Technologies Limited, Beijing 100193, China. \\
}

\address[yangshuaiaddress]{Department of Mathematics, University of British Columbia, Vancouver V6T1Z2, Canada.}

\address[sjtuaddress]{School of Mathematical Sciences, Institute of Natural Sciences and MOE-LSC, Shanghai Jiao Tong University, Shanghai 200240, China.}

\begin{abstract}
This paper introduces the {\tt MeshAC} package, which generates three-dimensional adaptive meshes tailored for the efficient and robust implementation of multiscale coupling methods. While Delaunay triangulation is commonly used for mesh generation across the entire computational domain, generating meshes for multiscale coupling methods is more challenging due to intrinsic discrete structures such as defects, and the need to match these structures to the continuum domain at the interface. The {\tt MeshAC} package tackles these challenges by generating meshes that align with fine-level discrete structures. It also incorporates localized modification and reconstruction operations specifically designed for interfaces. These enhancements improve both the implementation efficiency and the quality of the coupled mesh. Furthermore, {\tt MeshAC} introduces a novel adaptive feature that utilizes gradient-based a posteriori error estimation, which automatically adjusts the atomistic region and continuum mesh, ensuring an optimal balance between accuracy and efficiency. This package can be directly applied to the geometry optimization problems of a/c coupling in static mechanics~\cite{2006_SP_PB_TO_IJMCE, tembhekar2017automatic, hodapp2019lattice, 2014-bqce, LuOr:acta}, with potential extensions to many other scenarios. Its capabilities are demonstrated for complex material defects, including straight edge dislocation in BCC W and double voids in FCC Cu. These results suggest that {\tt MeshAC} can be a valuable tool for researchers and practitioners in computational mechanics.






\end{abstract}


\end{frontmatter}

\input{notation.tex}

\section{Introduction}
\label{sec:intro}

Multiscale modeling and simulation address the limitations of single-scale models by combining different modeling paradigms to capture phenomena across multiple length and time scales. They have grained popularity across various fields, such as mechanical engineering, biology, materials science, and medical research ~\cite{weinan2011principles, horstemeyer2010multiscale, zeng2008multiscale, peng2021multiscale, tadmor2011modeling, fish2021mesoscopic}. Multiscale coupling methods~\cite{tembhekar2017automatic, Shenoy:1999a, tadmor1996quasicontinuum,fedosov2009triple,werder2005a,dupuis2007b,BEEX2014154} stand out as a typical class of multiscale computational approaches that aim to capture localized fine-scale features like material defects, through a domain decomposition strategy which employs fine-scale models in local regions of interest while adopts coarse-grained models in the far field. 

This paper aims to tackle the challenges related to mesh generation and adaptation in the atomistic/continuum (a/c) coupling, which serves as a representative multiscale coupling method. We provide {\tt MeshAC}, an open-source package specifically designed for addressing these challenges in physically relevant three-dimensional cases. The methodology developed here can be extended to more diverse scenarios, including discrete-to-continuum coupling~\cite{mikevs2017quasicontinuum}, random alloys ~\cite{beex2014multiscale}, truss structure~\cite{phlipot2019quasicontinuum}, polymer networks~\cite{ghareeb2020adaptive}, and magnetization dynamics~\cite{PhysRevResearch.2.013092}.



To that end, we first provide a brief introduction to the atomistic-to-continuum (a/c) coupling methods. Those methods have emerged as a promising approach to bridge the gap between atomistic and continuum models, enabling the simulation of large-scale crystalline solids with fine-scale resolution~\cite{tembhekar2017automatic, LuOr:acta, TadmorMiller:2012, van2020roadmap}. These methods have demonstrated significant potential for modeling crystalline defects, garnering attention from both engineering and mathematical communities. There are comprehensive overviews and benchmarks available for various a/c coupling methods in material defect simulation~\cite{van2020roadmap, miller2009unified}, and rigorous analyses of these methods have been discussed in detail~\cite{2014-bqce, LuOr:acta, luskin2013atomistic, colz2016, hodapp2021analysis}.

The efficient implementation of a/c coupling methods poses significant challenges, particularly in three dimensions, for the following reasons: 
\begin{itemize}
\item mesh generation must account for the presence of crystalline structures such as body-centered cubic (BCC) and face-centered cubic (FCC), as well as complex crystalline defects such as dislocations, voids, grain boundaries, and cracks. This leads to anisotropic atomistic regions and, consequently, anisotropic computational domains. 
\item 
ensuring effective a/c coupling requires the mesh to conform to the underlying atomistic grid, particularly at the a/c interfaces. As the interface evolves, it becomes necessary to employ sophisticated mesh operations at these interfaces, striking a balance between maintaining accuracy and minimizing computational costs.
\item 
to achieve an optimal trade-off between cost and accuracy, it becomes essential to utilize robust and efficient a posteriori estimators. These estimators play a crucial role in driving the relocation of the a/c interface, computational domain boundaries, and the redistribution of the mesh. For example, atomistic interface and computational boundary could potentially be updated by interface motion driven by error distribution as demonstrated in \cite{adapqmmm2022}.
\end{itemize}
The mesh generation and adaptation play a crucial role in achieving an optimal distribution of atomistic and continuum degrees of freedom, as well as ensuring an appropriate error distribution. This process significantly influences the performance of relaxation algorithms and, consequently, the successful implementation of a/c coupling methods for simulating material defects.


Currently, the majority of publicly available a/c coupling codes are limited to two-dimensional contexts. Examples include the Fortran-based code developed by Tadmor et al.~\cite{qc3D} and the concurrent atomistic-continuum (CAC) simulation framework~\cite{xu_payne_chen_liu_xiong_chen_mcdowell_2018}. It is worth noting that these codes do not prioritize mesh generation and adaptation. Codes for 3D a/c coupling simulations are primarily found within academic research groups and are not publicly available~\cite{ariza2012hotqc, marian2009finite, Kochmann_codes}. However, it is important to highlight that the aforementioned research has not specifically focused on the adaptive computations of 3D a/c coupling methods, which hold significant physical significance but remain unexplored to the best of the authors' knowledge. Relevant studies related to 2D simulations can be found in~\cite{tembhekar2017automatic, wang2018posteriori}.


In this paper, we introduce {\tt MeshAC}, a comprehensive 3D mesh generation and adaptation package specifically designed for a/c coupling simulations of crystalline defects. {\tt MeshAC} includes a mesh generation tool based on the {\tt Tetgen} package~\cite{si2015tetgen} and a customized mesh adaptation tool for the evolution of the atomistic/continuum interface.  The target users are engineering and scientific professionals involved in geometric optimization problems of static mechanics. {\tt MeshAC} offers the following features that effectively address the challenges associated with the efficient implementation of three-dimensional a/c coupling methods, and enable accurate and robust modeling of complex crystalline defects.

\begin{itemize}
\item The mesh generation capabilities of the {\tt MeshAC} package are designed to respect the crystalline structure and the configuration of defects.
\item The {\tt MeshAC} package incorporates mesh adaptation capabilities that involve both atomistic mesh extension and local mesh refinement in the continuum region. The process of obtaining the coupled mesh involves a two-step procedure. Firstly, the mesh modification techniques are applied to the continuum region, resulting in mesh refinement. Subsequently, the mesh reconstruction techniques are employed to establish the new atomistic region. This two-step approach ensures the accurate and efficient generation of the coupled mesh.
\item During the simulation, {\tt MeshAC} automatically extends the atomistic region and adapts the finite element mesh to achieve a (quasi-)optimal balance between accuracy and efficiency, guided by a heuristic gradient-based {\it a posteriori} error estimator. 
\end{itemize}

The paper demonstrates the capabilities of the {\tt MeshAC} package by applying the BGFC method (atomistic/continuum blending with ghost force correction)~\cite{colz2016} to practical material defects, including straight edge dislocations and double voids. Furthermore, the paper showcases the potential extensions of the method to other types of defects, such as dislocation loops and grain boundaries. These examples highlight the versatility and effectiveness of the {\tt MeshAC} package in modeling and analyzing a wide range of material defects. These results indicate that {\tt MeshAC} can be a valuable tool for researchers and practitioners in the field of computational mechanics who are interested in studying the behavior of materials with complex defects.


\textbf{Organization}: The paper is organized as follows:
Section~\ref{sec:method} provides a detailed description of the methodology implemented in {\tt MeshAC}. Section~\ref{sec:numer} presents  numerical examples that demonstrate the application of {\tt MeshAC} to simulate crystalline defects. Section~\ref{sec:con} concludes the article with a discussion of the observations and outlines potential future work. 
Auxiliary results are presented in Appendix~\ref{sec:app:A}.

\section{Method}
\label{sec:method}

The accuracy and efficiency of a/c coupling simulations for material defects heavily rely on the quality of the generated mesh and its implementation. Hence, this section provides a detailed introduction to three-dimensional mesh generation and adaptation techniques that are specifically tailored for a/c coupling methods. The aim is to enhance the overall performance and reliability of simulations in this context.

\begin{figure}[htb]
	\centering 
	\includegraphics[height=7cm]{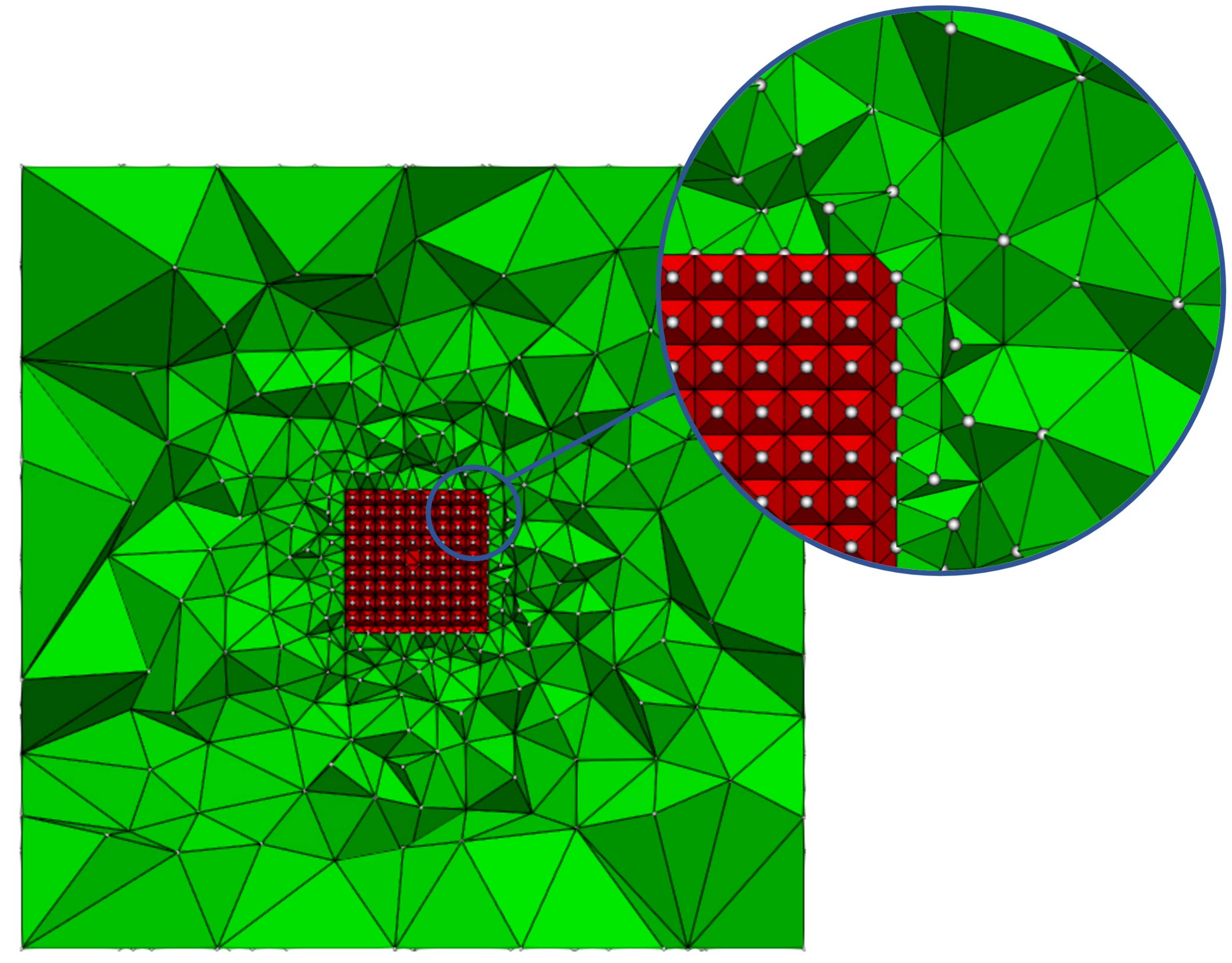}
	\caption{The figure depicts a 2D slice of the mesh $\T_h$, where the atomistic region $\T^{\rm a}$ is shown in red, and the continuum region $\T^\c$ is shown in green. White dots in the red region represent the atoms located within $\Omega^{\rm a}$, and mesh nodes in the continuum region $\Omega^{\rm c}$.}
	\label{fig:acmesh}
\end{figure}

To illustrate the a/c coupling model, we divide the computational domain $\Omega \subset \mathbb{R}^3$ into two sub-regions: the atomistic region and the continuum region. The tetrahedral mesh generated by the atomic lattice in the atomistic region $\Omega^{\rm a}$ is denoted as $\T^{\rm a}$, while the tetrahedral partition of the continuum region $\Omega^{\rm c}$ is referred to as $\T^\c$. We assume the presence of a sharp interface $\mathcal{I}$ between $\Omega^{\rm a}$ and $\Omega^{\rm c}$. The coupled tetrahedral mesh for a/c coupling methods is defined as $\T_h = \T^{\rm a} \cup \T^\c$, as depicted in Figure \ref{fig:acmesh}. The construction details of this coupled mesh will be presented in the subsequent part of this section.

In the context of static a/c coupling methods, the equilibrium displacement $u^{\rm ac}$ is determined by solving the geometry optimization problem $u^{\rm ac} \in \arg\min E_{\rm ac}(u)$, where the total coupling energy functional $E_{\rm ac}$ can be expressed as the sum of three components:
\[E_{\rm ac}(u) := E_{\rm a}(u) + E_{\rm c}(u) + E_{\rm corr}(u).\]
The energy functional $E_{\rm a}(u)$ and $E_{\rm c}(u)$ correspond to the atomistic and continuum regions, respectively. Additionally, $E_{\rm corr}(u)$ is the correction term utilized to eliminate spurious forces~\cite{LuOr:acta, COLZ2013}. Section~\ref{sec:sub:bgfc} will provide a detailed explanation of BGFC, a representative a/c coupling method implemented in {\tt MeshAC}.

\subsection{Mesh generation}
\label{sec:sub:mesh_gen}

\begin{figure}[H]
	\centering 
	\includegraphics[height=11cm]{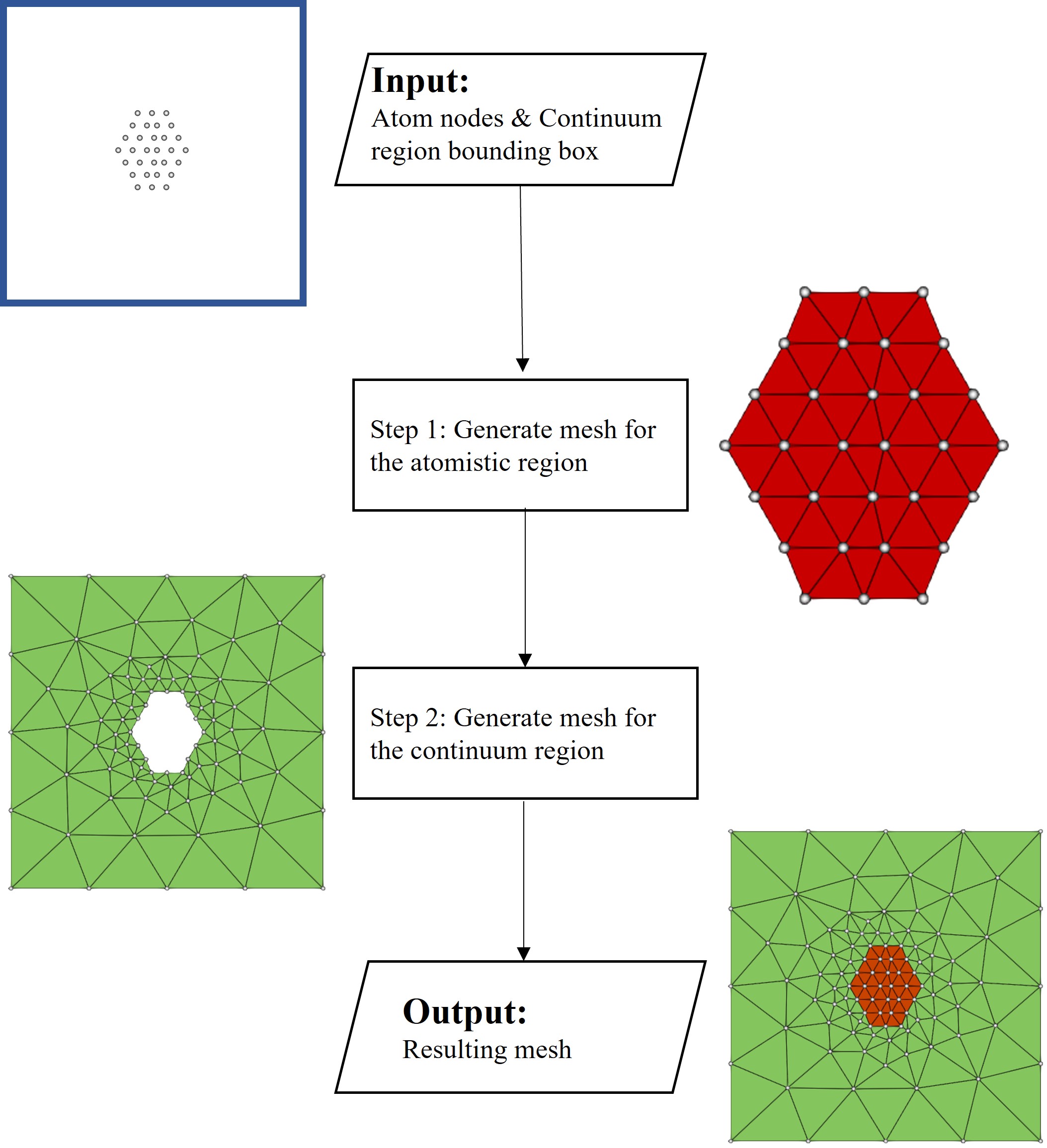}
	\caption{The workflow for the generation of the a/c coupling mesh $\T_h$ (2D slice). Firstly, given the atom positions and the computational domain $\Omega$, the canonical mesh $\T^{\rm a}$ for the atomistic region is generated using the techniques described in Section~\ref{sec:sub:sub:mesh_atom}. Next, the mesh $\T^{\rm c}$ for the continuum region is constructed using the methodology introduced in Section~\ref{sec:sub:sub:mesh_continum}. The combination of the meshes for both regions completes the process of establishing the coupled mesh $\T_h$.}
	\label{fig:meshgen}
\end{figure}

Unlike conventional approaches in finite element methods that often utilize Delaunay triangulation for mesh generation throughout the entire computational domain, the mesh generation approach proposed in this study is notably more intricate due to the presence of the atomistic region. Generating a three-dimensional coupled mesh $\T_h$ presents considerable challenges, as the atomistic region may lack convexity, and constructing the interface (surface) mesh necessitates careful consideration. To address these challenges, the {\tt MeshAC} package has been developed to effectively tackle the aforementioned issues. The workflow for constructing the coupled mesh $\T_h$ is presented in Figure~\ref{fig:meshgen}. Since $\T_h = \T^{\rm a} \cup \T^\c$, the meshes for different regions are generated separately, and then combined to form the coupled mesh. 

Specifically, given the atom positions and the computational domain $\Omega$, the first step is to generate the canonical mesh $\T^{\rm a}$ for the atomistic region $\Omega^{\rm a}$ using the techniques described in Section~\ref{sec:sub:sub:mesh_atom}. Next, the mesh $\T^{\rm c}$ for the continuum region $\Omega^{\rm c}$ is constructed using the methodology presented in Section~\ref{sec:sub:sub:mesh_continum}. By following these steps, the coupled mesh $\T_h$ is established.

In the subsequent sections, we will provide detailed explanations of each step involved in this process.

\subsubsection{Canonical mesh $\T^{\rm a}$ for the atomistic region}
\label{sec:sub:sub:mesh_atom}

\begin{figure}[H]
	\centering 
	\subfigure[]{
	\includegraphics[scale=0.5]{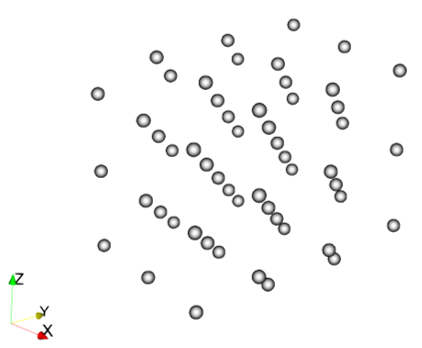}}
	\hskip 2cm
	\subfigure[]{
	\includegraphics[scale=0.5]{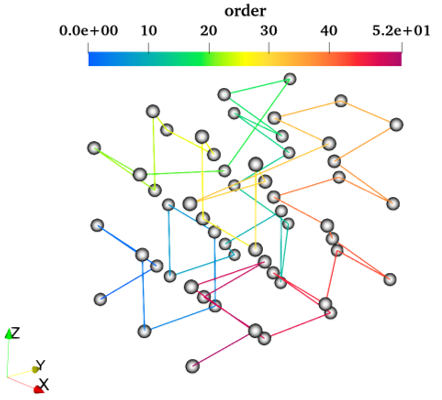}}
	\subfigure[]{
	\includegraphics[scale=0.5]{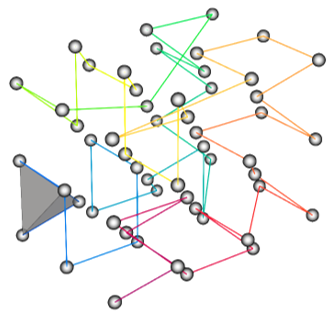}}
	\subfigure[]{
	\includegraphics[scale=0.5]{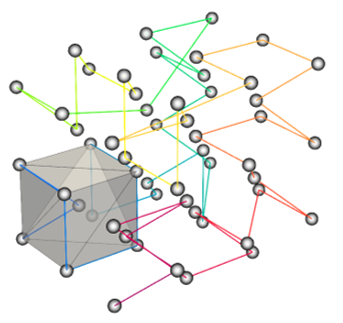}}
	\subfigure[]{
	\includegraphics[scale=0.5]{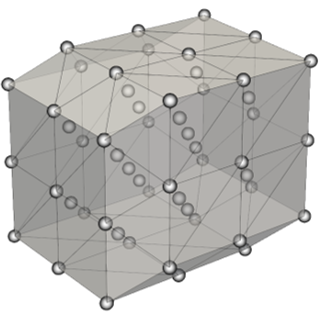}}
	\caption{The illustration of Algorithm~\ref{algorithm_delaunay_convex_hull} (the Delaunay triangulation of a 3-dimensional point set). (a) 3-dimensional point set; (b)The point set (i.e., atom positions) in $\R^{3}$ is sorted using the BRIO technique \cite{amenta2003incremental} to optimize the spatial search efficiency; (c) An initial tetrahedron element is then generated; (d) iterative application of the Bowyer-Watson algorithm~\cite{bowyer1981computing,watson1981computing} to construct the Delaunay triangulation incrementally; (e) the pre-processed mesh $\T^{\rm a-pre}$ is outputted.}
	\label{fig:point_triangulation}
\end{figure}

The initial step in generating the canonical mesh for the atomistic region involves utilizing the Delaunay triangulation method provided by the renowned {\tt Tetgen} package~\cite{si2015tetgen} on the set of atom positions, which can be treated as a point set in $\mathbb{R}^3$. This process yields a pre-processed mesh $\T^{\rm a-pre}$. Algorithm~\ref{algorithm_delaunay_convex_hull} outlines the iterative steps involved in constructing the triangulation by inserting one point at a time until the Delaunay criterion is satisfied. Figure~\ref{fig:point_triangulation} provides an illustration of the triangulation.

To enhance the efficiency of spatial search, the Biased Randomized Insertion Order (BRIO) technique~\cite{amenta2003incremental} is employed for sorting the atom positions. Additionally, to ensure the quality of the constructed elements in each iteration and guarantee a high-quality final result, the Bowyer-Watson algorithm~\cite{bowyer1981computing,watson1981computing} is utilized.

\begin{algorithm}[H]
	\caption{Delaunay triangulation of a 3D point set} 
	\label{algorithm_delaunay_convex_hull}
	\LinesNumbered  
	\KwIn{A point set $\mathtt{Q}$ in $\R^{3}$}
        $\mathtt{P}$ $\leftarrow$ Sort the points in $\mathtt{Q}$ using the BRIO technique\;
        $\mathcal{T}^{\rm a-pre}$ $\leftarrow$ Start with an initial tetrahedron element whose nodes are denoted as $\mathtt{E}$\;
        \ForEach{{\rm point} $p$ {\rm in} $\mathtt{P}\setminus \mathtt{E}$}{
            $\mathcal{T}^{\rm a-pre}$ $\leftarrow$ Apply the Bowyer-Watson algorithm to insert $p$ in $\mathcal{T}^{\rm a-pre}$\;
        }
        \KwOut{Pre-processed mesh $\mathcal{T}^{\rm a-pre}$}
\end{algorithm}

After constructing the pre-processed mesh $\T^{\rm a-pre}$ using the Delaunay triangulation algorithm, further manipulations are required to obtain the canonical mesh $\T^{\rm a}$ for the atomistic region. This is because the atomistic region may contain complex crystalline defects that result in anisotropic topology, requiring additional steps for refinement and adjustment.

Although the Delaunay triangulation algorithm can be used to generate a triangulation for the atomistic region after prescribing atomic degrees of freedom (DOFs). However, it is not appropriate to use the convex hull to identify surface atoms for the atomistic region. While Algorithm~\ref{algorithm_delaunay_convex_hull} is capable of generating a convex hull for a given set of points, it can potentially create elements of poor quality.

\begin{figure}[H]
	\centering 
	\subfigure[Pre-processed mesh $\T^{\rm a-pre}$]{
	\label{convex_hull}
	\includegraphics[scale=0.25]{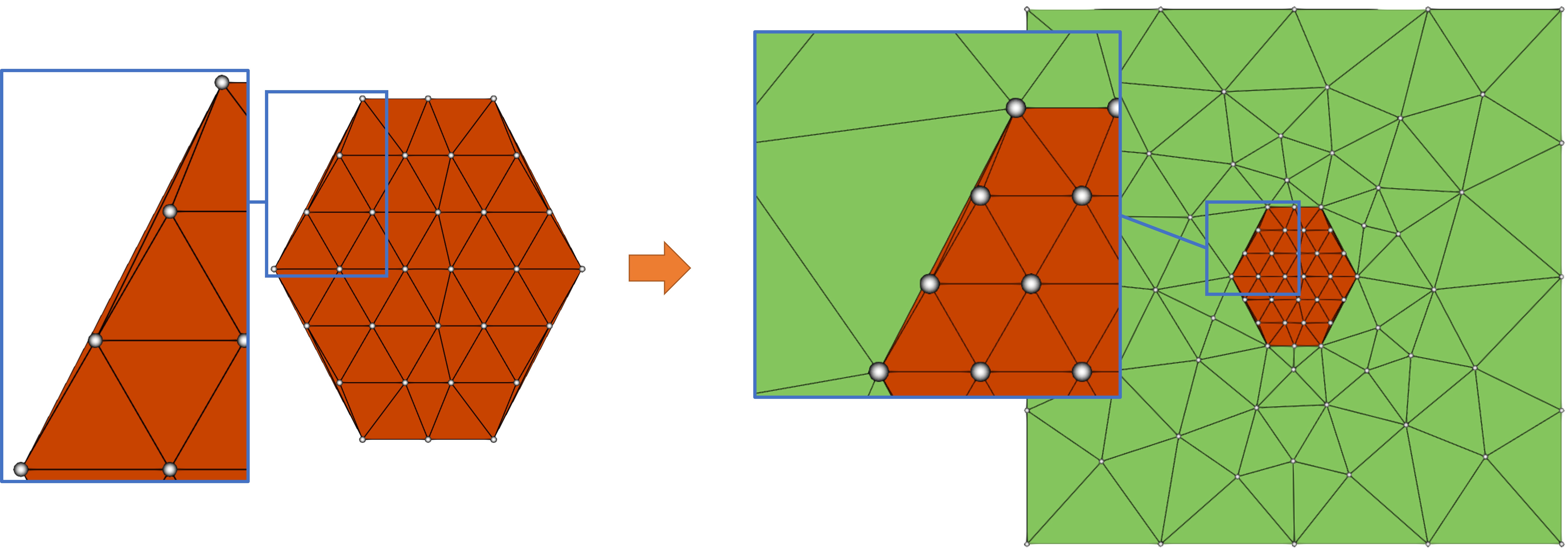}}
	\hskip 2cm
	\subfigure[Canonical mesh $\T^{\rm a}$]{
	\label{clean_interface}
	\includegraphics[scale=0.25]{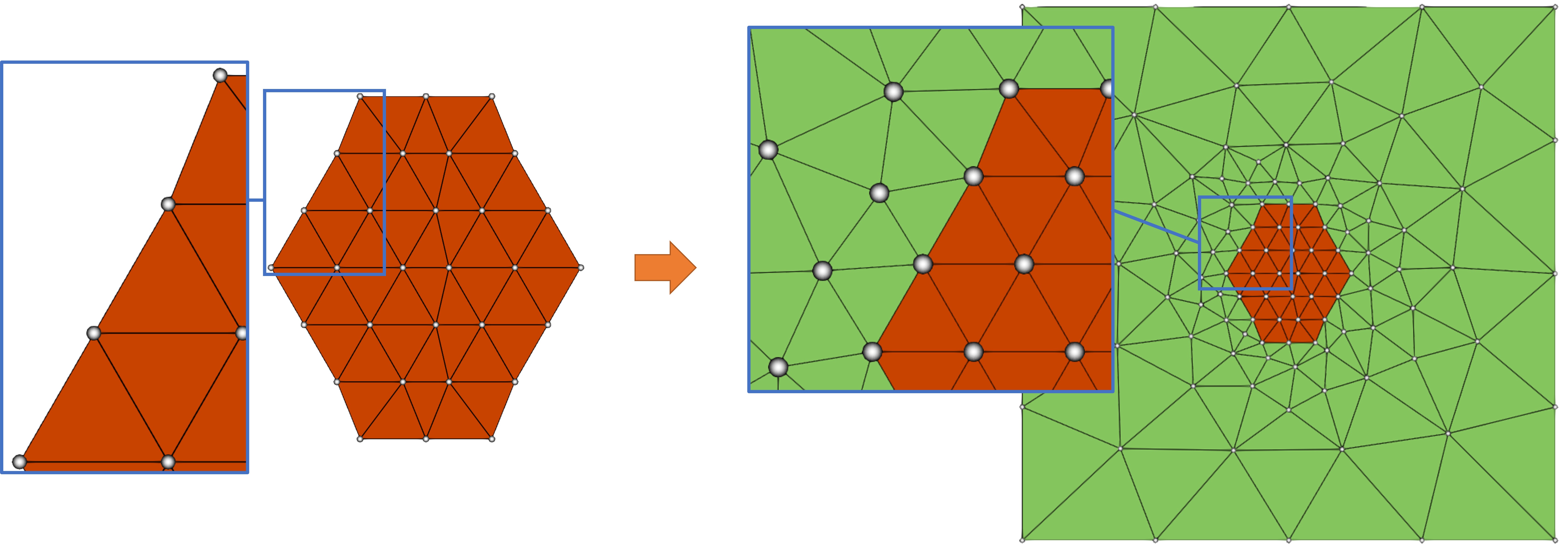}}
	\caption{The comparison between $\T^{\rm a-pre}$ and $\T^{\rm a}$, where the major difference is the interfaces of the atomistic region for edge dislocation (cf. Section~\ref{sec:sub:edge}), and the elements colored in green is the continuous mesh constructed in Section~\ref{sec:sub:sub:mesh_continum}.}
	\label{fig:atomistic_with_defects}
\end{figure}

There are two possible reasons for this. 
First, due to the irregular arrangement of surface atoms on the atomistic region caused by defects, it is challenging to define a convex hull precisely. Second, due to round-off errors, some atoms on the boundary of the hull (in exact arithmetic) may fall slightly inside or outside the convex hull, resulting in elements of poor quality. An example of the pre-processed mesh $\T^{\rm a-pre}$ for an edge dislocation is shown in Figure~\ref{convex_hull}.

To address these issues, a novel algorithm (Algorithm~\ref{algorithm_alpha_shape}) is introduced to identify and remove poor-quality elements, producing a mesh $\T^{\rm a}$ suitable for coupling, as illustrated in Figure~\ref{clean_interface}. This approach differs from mesh generation in traditional finite element methods and presents a unique challenge specific to the {\tt MeshAC} framework. The hyperparameter $r_{\rm max}$ in the algorithm can be set empirically or determined by finding the maximum distance between any two nearest atoms.

\begin{algorithm}[H]
	\caption{Element deleting algorithm} 
	\label{algorithm_alpha_shape}
	\LinesNumbered  
	\KwIn{The pre-processed mesh $\mathcal{T}^{\rm a-pre}$ and max radius $r_{\rm max}$}
        \KwOut{A canonical mesh $\T^{\rm a}$}
        Let $\T^{\rm a}=\mathcal{T}^{\rm a-pre}$\;
        \Repeat{$\mathcal{D}\neq\emptyset$}{
            Initialize two sets of elements $\mathcal{B}=\mathcal{D}=\emptyset$\;
            Put all the elements adjacent to the boundary of $\mathcal{T}^{\rm a}$ in $\mathcal{B}$\;
            \ForEach{{\rm element} $K$ {\rm in} $\mathcal{B}$}{
                \If{{\rm the bounded circle radius of} $K$ {\rm meets} $r>r_{\rm max}$}{
                    {\rm Add} $K$ {\rm in} $\mathcal{D}$\;
                }
            }
            $\mathcal{T}^{\rm a} \leftarrow \mathcal{T}^{\rm a} \setminus \mathcal{D}$\;
        }
\end{algorithm}

\subsubsection{Mesh Generation for the continuum region $\T^{\rm c}$} \label{sec:sub:sub:mesh_continum}

\begin{figure}[H]
	\centering 
	\subfigure[]{
	\includegraphics[scale=0.45]{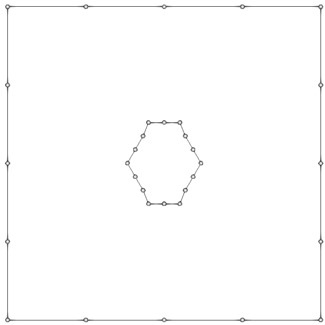}}
	\subfigure[]{
	\includegraphics[scale=0.45]{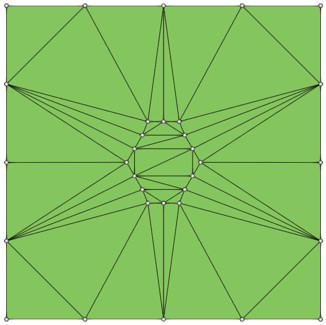}}
	\subfigure[]{
	\includegraphics[scale=0.45]{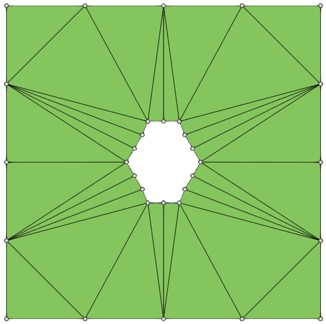}}
	\subfigure[]{
	\includegraphics[scale=0.45]{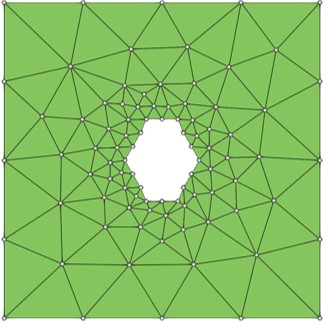}}
	\caption{The illustration of the continuum mesh generation (i.e., the construction of $\T^{\c}$) in a two-dimensional view. (a) Initialize the boundary mesh in continuum region and the boundary mesh in atomistic region; (b) Generate a Delaunay triangulation for all the given nodes; (c) Recover the boundary mesh and extract elements in atomistic region; (d) Refine the elements in continuum region. CDT: (a-c), QMR: (d)}
	\label{fig:continuum_region_mesh_gen_2D}
\end{figure}

In this section, we propose a novel approach to generate the continuum mesh $\T^{\rm c}$ that is compatible with the construction of the atomistic mesh $\T^{\rm a}$. This approach combines the constrained Delaunay triangulation (CDT)~\cite{si2005meshing,si20113d} and quality mesh refinement (QMR)~\cite{si2014incrementally} techniques. The procedure involves four main steps:

\begin{enumerate}[(a)]
\item Initialization of nodes at the boundary of the computational domain $\Omega$ and extraction of the boundary mesh from $\T^{\rm a}$.
\item Generation of a tentative Delaunay triangulation for all the given nodes, which includes some invalid elements within the atomistic region.
\item Recovery of the boundary mesh and removal of invalid elements within the atomistic region using the algorithm proposed in \cite{si20113d}.
\item Refinement of elements in the continuum region using the methodology described in \cite{si2014incrementally}.
\end{enumerate}
This approach ensures that the elements in $\T^{\rm c}$ achieve optimal quality while maintaining a smooth transition of mesh size in the resulting coupled mesh $\T_h$. Figure~\ref{fig:continuum_region_mesh_gen_2D} provides an illustration of the generated continuum region mesh $\T^{\rm c}$.

\subsection{Mesh adaptation}
\label{sec:sub:mesh_adap}

In simulations involving atomistic/continuum coupling, a significant challenge is to achieve an optimal assignment of the atomistic and continuum regions, along with an appropriate mesh structure, that strikes a balance between accuracy and efficiency~\cite{wang2018posteriori, wang2021posteriori, fu2023adaptive}. Hence, the use of a robust and effective mesh adaptation technique becomes crucial in such simulations.

Adaptive atomistic/continuum (a/c) coupling methods introduce additional challenges in automatically partitioning the computational mesh and dynamically adjusting the model, compared to classical adaptive finite element methods. These challenges are particularly pronounced in 3D simulations where the interface between the atomistic and continuum regions needs to be updated during the adaptive process. The selection of an appropriate adaptive algorithm is closely intertwined with the choice of the a/c coupling method and the construction of the {\it a posteriori} error estimator~\cite{wang2018posteriori}. Therefore, a detailed discussion of the adaptive algorithm is deferred to Sections~\ref{sec:sub:bgfc} and \ref{sec:app:A}.

\begin{figure}[H]
	\centering 
	\includegraphics[height=12cm]{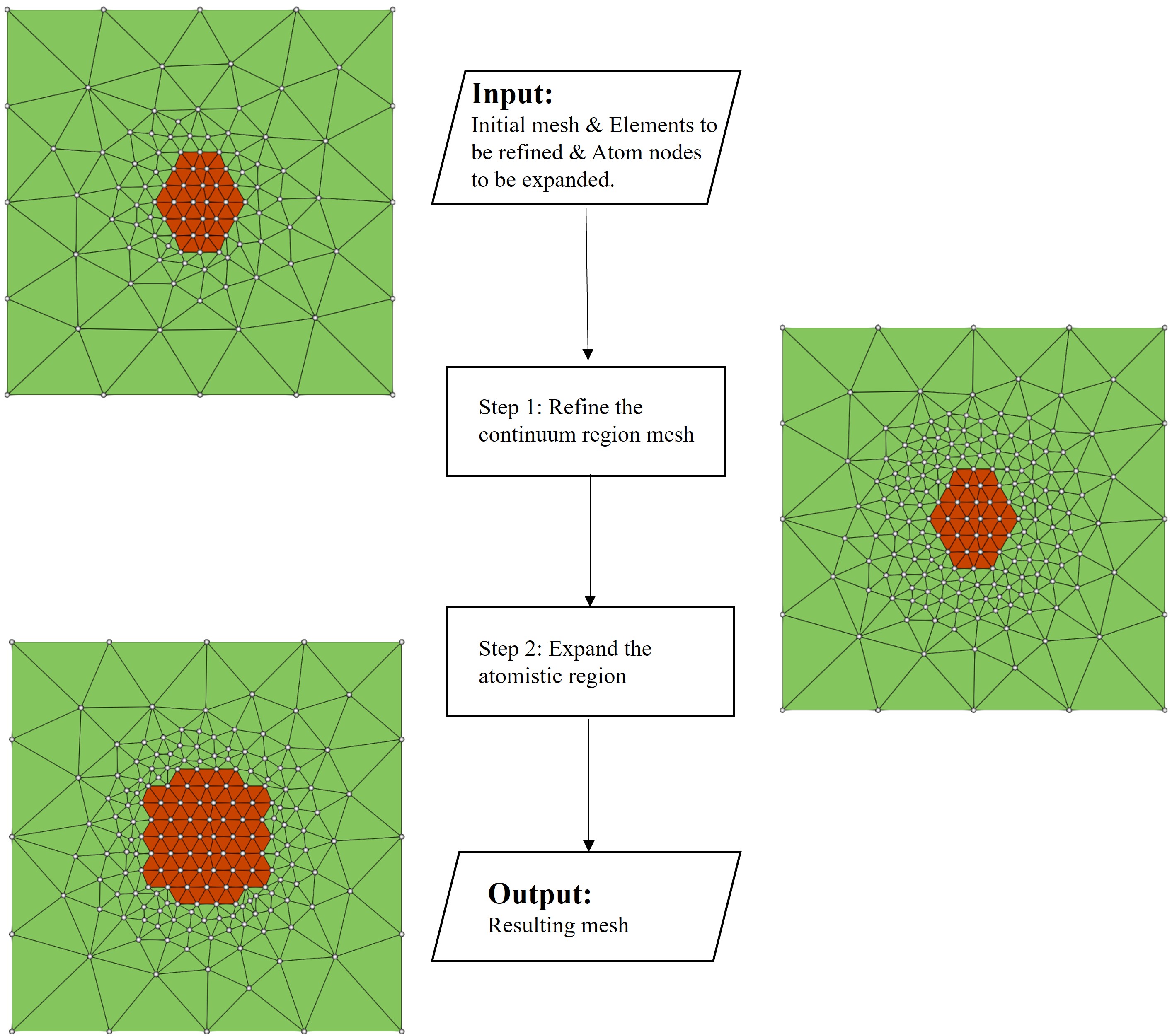}
    	\caption{The workflow of the mesh adaptation process (2D slice) from $\T_h$ to $\T_h^{\rm new}$. The procedure is divided into two steps, where we first refine $\T^{\rm c}_{\rm ref}$ in the continuum region based on the mesh modification techniques introduced in Section~\ref{sec:sub:sub:refine_c}, then establish the new atomistic region using the mesh reconstruction techniques given in Section~\ref{sec:sub:sub:ext}.}
	\label{fig:meshadapt}
\end{figure}

In this section, our focus is on the implementation of mesh adaptation in {\tt MeshAC}, which includes extending the atomistic mesh and performing local mesh refinement in the continuum region. We assume that the adaptive algorithm has already provided a set of elements that need to be refined or extended (cf. Algorithm~\ref{alg:refine}). The mesh adaptation procedure takes the following inputs: (1) the current coupled mesh $\T_h$ that requires adjustment, (2) a set of elements to be refined in the continuum region denoted as $\T^{\rm c}_{\rm ref}$, and (3) a set of canonical meshes $\T^{\rm a}_{\rm ext}$ that need to be extended. The procedure to obtain the new coupled mesh $\T_h^{\rm new}$ consists of two steps. First, we refine $\T^{\rm c}_{\rm ref}$ in the continuum region using the mesh modification techniques introduced in Section~\ref{sec:sub:sub:refine_c}. Then, we establish the new atomistic region using the mesh reconstruction techniques described in Section~\ref{sec:sub:sub:ext}. The workflow is illustrated in Figure~\ref{fig:meshadapt}.

\subsubsection{Local mesh refinement in the continuum region}
\label{sec:sub:sub:refine_c}

\begin{figure}[H]
	\centering 
	\includegraphics[height=4cm]{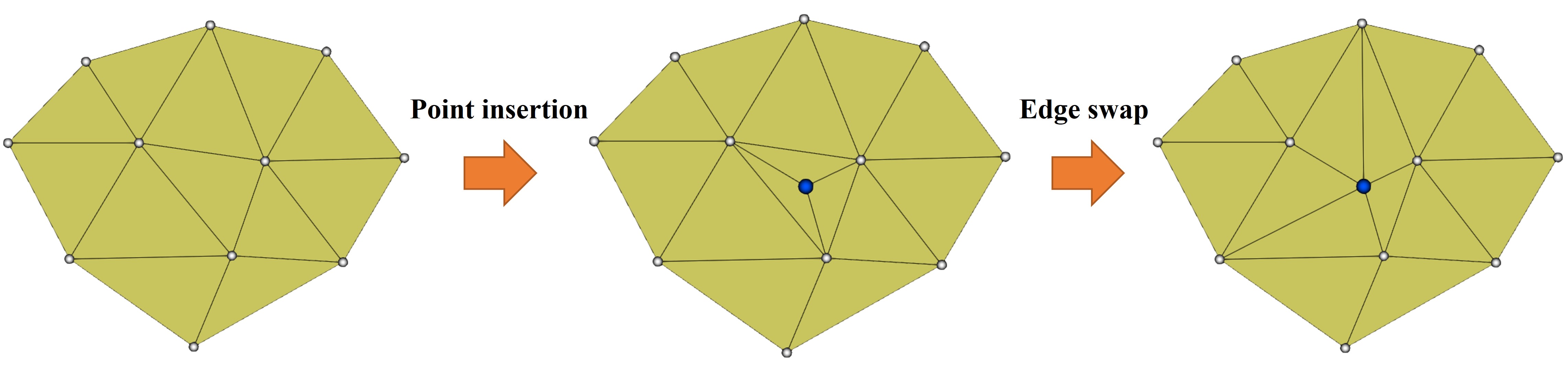}
	\caption{Illustration of local mesh refinement in the continuum region.}
	\label{fig:mesh_adapt_refine}
\end{figure}

In contrast to the mesh refinement technique used in generating $\T^{\rm c}$ as discussed in Section~\ref{sec:sub:mesh_gen}, the refinement process employed here is controlled by an {\it a posteriori} error estimator. A practical example of this error estimator will be presented in Section~\ref{sec:sub:bgfc}. Assuming that $\T^{\rm c}_{\rm ref}$ has already been obtained for refinement, two techniques are utilized in {\tt MeshAC}: the edge swap operation and point insertion. These techniques are illustrated in Figure~\ref{fig:mesh_adapt_refine}.

\begin{figure}[H]
	\centering 
	\includegraphics[scale=0.6]{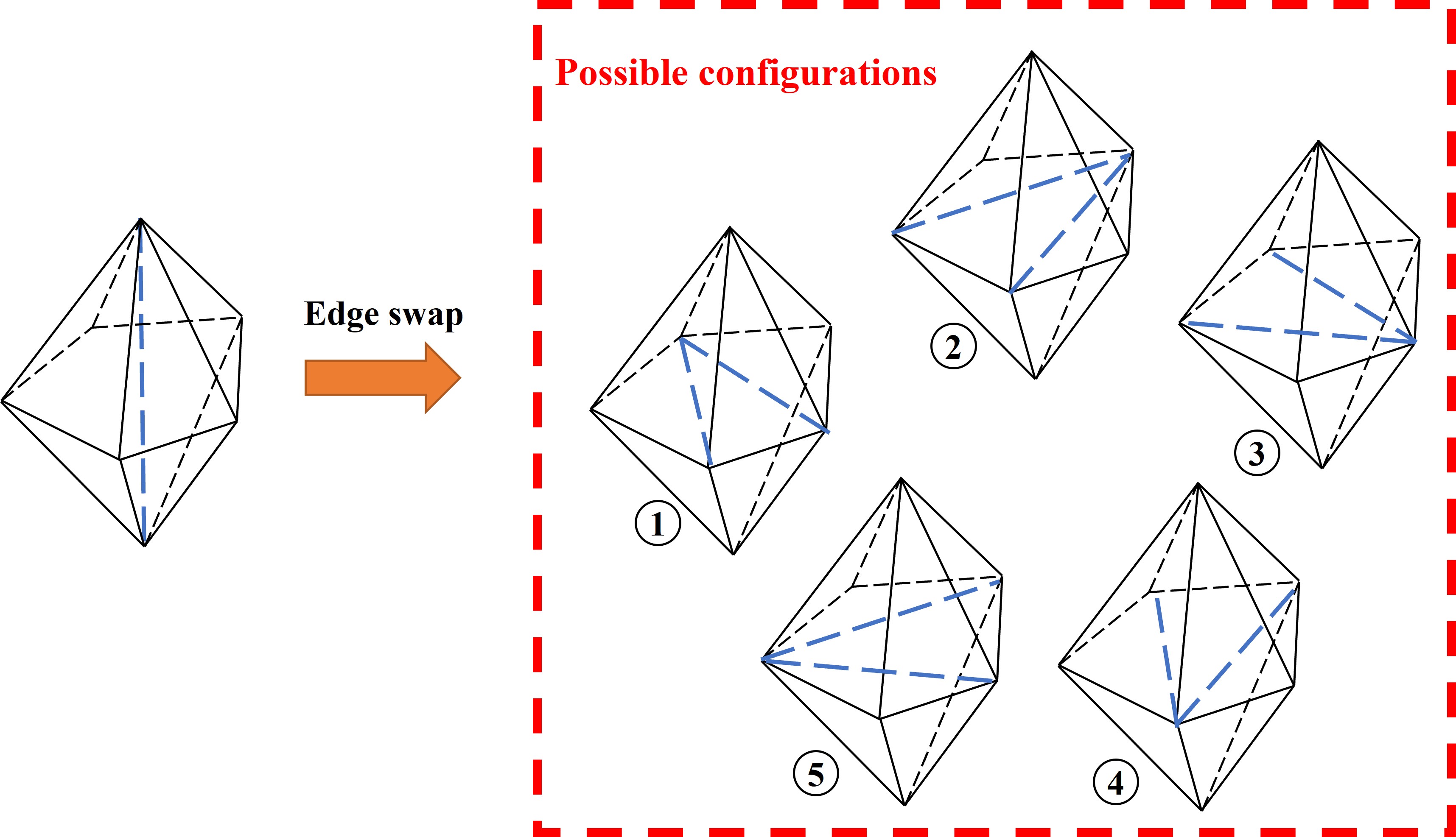}
	\caption{The illustration of multiple potential configurations for an edge swap operation.}
	\label{fig:edgeswap}
\end{figure}

The edge swap operation is a technique used to improve the local quality of elements by reestablishing connections among nodes adjacent to an edge. Typically, there are multiple potential configurations for an edge swap operation, as illustrated in Figure~\ref{fig:edgeswap}. Before implementing an edge swap, the tetrahedral quality of each configuration needs to be evaluated. The quality of a tetrahedron $K$ is defined as follows:
\begin{equation}\label{eq:element_quality}
		q(K)=\frac{72\sqrt{3} \vert K \vert }{(\sum_{i = 1}^{6} s(e_{i})^2)^{\frac{3}{2}}  }\in[0,1],
\end{equation}
where $\vert K \vert$ represents the volume of tetrahedron $K$, and $s(e_i)$ is the length of the $i$-th edge of $K$. It should be noted that $q(K)=1$ if $K$ is a regular tetrahedron. Furthermore, we define the quality of a tetrahedral element set $\mathcal{K}=\{K_{i}\}$ as
\begin{equation}
    Q(\mathcal{K})=\min_{K_i \in \mathcal{K}}{q(K_{i})}.
\end{equation}

Let $\mathcal{A}$ be the collection of tetrahedral elements that share an edge to be swapped, and let $\mathcal{B}$ represent the collection of all possible tetrahedral configurations resulting from the edge swap operation. To determine the acceptability of a configuration in $\mathcal{B}$, we apply the following condition: $1.01Q(\mathcal{A})<Q(\mathcal{B})$, where $Q(\cdot)$ denotes the tetrahedral quality function.

To ensure that the resulting mesh is of high quality, it is crucial to rigorously validate the feasibility of each potential configuration in $\mathcal{B}$. This validation involves conducting a series of topological tests to confirm that the proposed configuration does not violate any geometric constraints. The feasibility of a particular configuration is determined by examining the conditions necessary for it to be valid. These conditions include ensuring the non-intersection of edges, the absence of triangle flipping, and the preservation of a consistent orientation for adjacent faces. Only configurations that are both feasible and meet the quality criteria are considered acceptable for implementation.

In summary, we introduce Algorithm~\ref{algorithm_refine} to describe the refinement process. This algorithm involves handling the elements marked for refinement in batches, dividing each element into four sub-elements, and utilizing edge swap operations to improve the local mesh quality. The objective of this algorithm is to optimize the distribution of the computational mesh, striking a balance between computational efficiency and accuracy in atomistic/continuum coupling simulations.

\begin{algorithm}[H]
	\caption{Local mesh refinement in continuum region} 
	\label{algorithm_refine}
	\LinesNumbered  
	\KwIn{$\T^{\rm c}$ and $\mathcal{T}^{\rm c}_{\rm ref}$\;
	}
        Initialize an element set $\mathcal{X}=\emptyset$\;
	\ForEach{{\rm element} $K$ {\rm in} $\mathcal{T}^{\rm c}_{\rm ref}$}{
            Split $K$ into four tetrahedral by inserting a node in the barycenter of $K$\;
            Update $\T^{\rm c}$ and add the new elements in $\mathcal{X}$\;
	}
        \Repeat{$\mathcal{X}=\emptyset$}{
            Initialize an element set $\mathcal{Y}=\emptyset$\;
            \ForEach{{\rm element} $K$ {\rm in} $\mathcal{X}$}{
                \If{K exists}{
                    \ForEach{{\rm edge} $e$ {\rm in} $K$}{
                        \If{$e$ {\rm does not belong to the boundary of atomistic region as well as meets the swap condition}}{
                            Apply edge swap operation for $e$\;
                            Update $\mathcal{T}^{c}$ and add the new elements in $\mathcal{Y}$\;
                            break\;
                        }
                    }
                }
            }
           $\mathcal{X}=\mathcal{Y}$ \;
        }
		
\end{algorithm}

\subsubsection{The extension of atomistic region}
\label{sec:sub:sub:ext}

In the subsequent stage of the mesh adaptation process, we move on to constructing a new coupled mesh $\T^{\rm new}_h$ by extending the atomistic region. The algorithmic details for this process are presented in Algorithm~\ref{algorithm_expand}, which we describe as follows.

\begin{figure}[H]
	\centering 
	\subfigure[\label{ext-a-1}]{
	\includegraphics[scale=0.5]{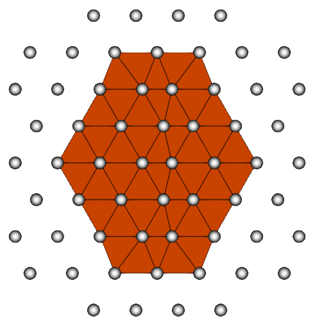}}
	\subfigure[\label{ext-a-2}]{
	\includegraphics[scale=0.5]{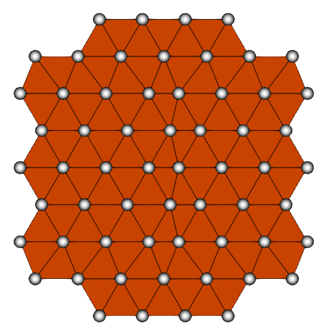}}
	\subfigure[\label{ext-a-3}]{
	\includegraphics[scale=0.5]{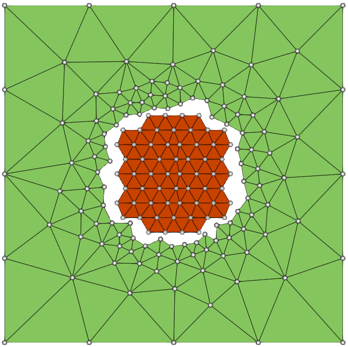}}
	\subfigure[\label{ext-a-4}]{
	\includegraphics[scale=0.5]{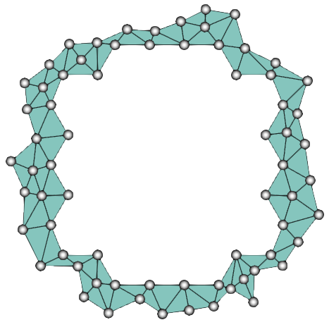}}
	\subfigure[\label{ext-a-5}]{
	\includegraphics[scale=0.5]{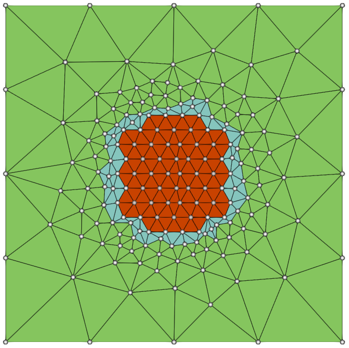}}
 	\subfigure[\label{ext-a-6}]{
	\includegraphics[scale=0.5]{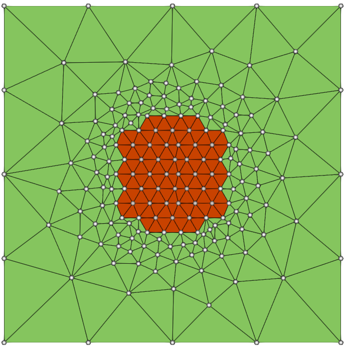}}
	\caption{The illustration of atomistic region extension and the construction of new coupled mesh $\T_h$. The procedure contains: (a) begin with the original atomistic mesh and the marked atoms to be involved in; (b)
 mesh reconstruction in the new atomistic region; (c) remove all the elements in the continuum mesh that are intersected with the new atomistic region; (d) construct a temporary sub-mesh for the cavity; (e) fuse all the sub-meshes; (f) new coupled mesh $\T_h$ for next-step simulations.}
	\label{fig:atom_expend}
\end{figure}

Initially, the atoms to be added to the new atomistic region are determined by implementing the marking step in Algorithm~\ref{alg:refine}. Subsequently, the atomistic mesh is reconstructed by combining these marked atoms with the existing atoms from the original atomistic region, utilizing the technique outlined in Section~\ref{sec:sub:sub:mesh_atom}. As the mesh for the new atomistic region intersects with the continuum region, it is crucial to remove any continuum elements that intersect with the new atomistic region. This results in the creation of a cavity layer, as illustrated in Figure~\ref{ext-a-4}, with boundaries formed by both the  continuum region (after removing the elements intersecting with the new atomistic region) and the new atomistic region.

Next, a temporary sub-mesh is constructed for the cavity using the method discussed in Section~\ref{sec:sub:sub:mesh_continum}. Finally, the resulting mesh is obtained by merging all the sub-meshes together. To ensure a smooth transition of the fused mesh elements, Laplacian smoothing \cite{herrmann1976laplacian} is applied to the continuum mesh nodes around the fusion region.

This entire procedure is depicted in Figure~\ref{fig:atom_expend}, which bears a resemblance to the approach presented in \cite[Figure 3]{wang2018posteriori} for two-dimensional problems.

\begin{algorithm}[H]
	\caption{Atomistic region extension and new coupled mesh construction} 
	\label{algorithm_expand}
	\LinesNumbered  
	\KwIn{$\mathcal{T}^{\rm c}$ generated by Algorithm~\ref{algorithm_refine} and a set of atom $\mathtt{P}$ marked by Algorithm~\ref{alg:refine};}
 	\tcp{Reconstruct the atomistic region mesh}
        Extract the original atom nodes in set $\mathtt{Q}$, let $\mathtt{P} = \mathtt{P} \cup \mathtt{Q}$\;
        Generate a new atomistic mesh $\mathcal{T}^{\rm a}$ based on $\mathtt{P}$ by applying the method introduced in Section \ref{sec:sub:sub:mesh_atom}\;
        \tcp{Remove elements in the original continuum region mesh}
        Remove the elements in $\mathcal{T}^{\rm c}$ that intersect with $\mathcal{T}^{\rm a}$\;
        \tcp{Construct a temporary sub-mesh}
        Extract the inner boundary mesh of $\mathcal{T}^{\rm c}$ as $\mathcal{S}$ and outer boundary mesh of $\mathcal{T}^{\rm a}$ as $\mathcal{S}^{\rm a}$\;
        Generate a temporary sub-mesh mesh $\mathcal{T}^{\rm tmp}$ based on $\mathcal{S}$ and $\mathcal{S}^{\rm a}$  by exploiting the method given in Section \ref{sec:sub:sub:mesh_continum}\;
        \tcp{Construct the coupled mesh by fusing the meshes}
        Generate the new coupled mesh $\mathcal{T}^{\rm new}_h$ by fusing  $\mathcal{T}^{\rm c}$, $\mathcal{T}^{\rm a}$, and $\mathcal{T}^{\rm tmp}$\;
\end{algorithm}

\subsection{Spatial search tree based fast interpolation}

In practical simulations, the use of an efficient linear interpolation method is crucial. An inefficient interpolation method can substantially increase the computational cost and compromise the overall accuracy of the simulation. Therefore, careful attention is given to selecting and implementing an efficient interpolation method within {\tt MeshAC}. The interpolation tool is primarily designed to provide a reliable initial estimate for the atomistic geometry optimization problem (cf.~\eqref{eq:variational-A-problem}), as discussed in detail in our recent work~\cite{fu2023adaptive}. During the adaptive computations, the interpolation is utilized to interpolate values from the original coupled mesh $\T_h$ to the new coupled mesh $\T_h^{\rm new}$.

To expedite the interpolation procedure, we employ spatial search trees, as outlined in Algorithm~\ref{algorithm_interpolate}. The fundamental approach involves constructing a Kd-tree for all nodes in the initial mesh and an AABB-tree for all tetrahedral elements in the initial mesh. We then iterate through the nodes in the adapted mesh. If a node coincides with a node in the initial mesh (based on the Kd-tree), the corresponding solution value is directly retrieved. Otherwise, we locate a base element containing the node using the AABB-tree. The implementation of the Kd-tree and AABB-tree algorithms in {\tt MeshAC} is based on \cite{gitKdtree} and \cite{gitAABB}, respectively. 

Suppose the number of nodes in $\T_h^{\rm new}$ is $N$, and the number of nodes and tetrahedral elements in $\T_h$ are $M_{0}$ and $M_{1}$, respectively. The time complexity of the interpolation algorithm is $O(N\cdot(\log{M_{0}}+\log{M_{1}}))$, while the time complexity of the brute-force interpolation algorithm is $O(N\cdot(M_{0}+ M_{1}))$. By utilizing spatial search trees, we achieve a more efficient interpolation process compared to the brute-force method.


\begin{algorithm}[H]
	\caption{Interpolation from $\T_h$ to $\T_h^{\rm new}$} 
	\label{algorithm_interpolate}
	\LinesNumbered  
	\KwIn{The original coupled mesh $\mathcal{T}_h$ with nodal values and the new coupled mesh $\mathcal{T}_h^{\rm new}$ generated by Algorithm~\ref{algorithm_expand} \;}
	\KwData{
		$\epsilon$: a predefined tolerance\;
	}
    Build a Kd-tree $\mathsf{T}_{\rm kd}$ for all nodes in $\mathcal{T}_{h}$\;
    Build a AABB-tree $\mathsf{T}_{\rm aabb}$ for all tetrahedral elements in $\mathcal{T}_{h}$\;
	\ForEach{{\rm node} $p$ {\rm in} $\mathcal{T}^{\rm new}_h$}{
	    Search a node $q$ in a radius of $\epsilon$ from the $p$ in the $\mathsf{T}_{\rm kd}$\;
	    \eIf{$q$ {\rm exists}}{
     	    \tcp{$p$.value means the solution value of $p$}
	        $p$.value = $q$.value\;
	    }{
	    	Search a tetrahedra $K$ in the $\mathsf{T}_{\rm aabb}$ that contains $p$ with a tolerance $\epsilon$ \;
	    	$p$.value = 0\;
	    	\ForEach{{\rm endpoint} $q_{i}$ {\rm in} $K$}{
	    	\tcp{$w_i$ is the i-th barycentric coordinate of $p$ in $K$ }
	    	    $p$.value = $p$.value + $w_i\cdot q_{i}$.value\;
	    	}   	
	    }
	}		
\end{algorithm}

\section{Numerics}
\label{sec:numer}

This section showcases the capabilities of {\tt MeshAC} in solving a range of atomistic mechanics problems. Before delving into the demonstrations, we first highlight one specific a/c coupling method implemented in {\tt MeshAC}. While our package is versatile and compatible with various a/c coupling methods, we have chosen to adopt the blended ghost force correction (BGFC) method~\cite{fang2020blended, colz2016}. This method achieves the highest possible convergence rate when Cauchy-Born continuum model is used in the continuum region, as rigorously proven in ~\cite{colz2016}. By utilizing the BGFC method, we can attain superior performance and accuracy in our simulations.

All numerical tests we report in this section were conducted on a computer with an {\tt Intel(R) Core(TM) i5-7820HQ CPU @2.90GHz} and {\tt macOS (x86-64-apple-darwin19.6.0)} system.

\subsection{Atomistic model and BGFC method}
\label{sec:sub:bgfc}

\def\Rcore{R_{\rm DEF}}
\def\UsH{{\mathscr{U}}^{1,2}}
\def\Adm{{\rm Adm}}
\def\Use{\Us^{1,2}}

We first present the atomistic model as the reference model, along with the corresponding blended ghost force correction (BGFC) method. To simplify the presentation, we will omit certain technical details and provide them  in~\ref{sec:app:A}.

\subsubsection{Atomistic model}
\label{sec:sub:sub:A}
Given a non-singular matrix $\mA\in\R^{3\times 3}$, let $\Lhom=\mA\Z^3$ and $\L\subset \R^3$ represent a perfect single lattice possessing no defects, and the corresponding single lattice with some defects, respectively. The mismatch between $\L$ and $\Lhom$ characterizes possible defects such as point defects, dislocations and cracks. Let $\Us := \{v: \L\to \mathbb{R}^3 \}$ be the set of vector-valued lattice functions and $V_{\ell}$ be the site potential representing the energy distributed to the atomic site $\ell\in\Lambda$, which satisfies some fundamental assumptions on regularity and symmetry~\cite[Section 2]{2013-defects}. The energy-difference functional of the atomistic model is then defined by
\begin{align}
\label{energy-difference}
\E(u) =& \sum_{\ell\in\Lambda}\Big(V_{\ell}\big(Du_0(\ell)+Du(\ell)\big)-V_{\ell}\big(Du_0(\ell)\big)\Big),
\quad
\end{align}
where $D$ represents the finite-difference stencil operator and $u_0$ is a far-field {\it predictor} enforcing the presence of defects. The concrete definitions will be given in~\ref{sec:app:A}.

The equilibrium is obtained by solving the following geometry optimization problem
\begin{eqnarray}\label{eq:variational-A-problem}
u^{\rm a} \in \arg \min \big\{\E(u) ~\big|~ u \in \Use \big\}.
\end{eqnarray}
where ``$\arg\min$'' is understood as the set of local minima and the functional space of finite-energy displacements $\Use$ is defined in~\ref{sec:app:A} (cf.~\eqref{eq:spce-a}).

\subsubsection{BGFC method}
\label{sec:sub:sub:bgfc}

We decompose $\Omega$ into three regions: the {{\it atomistic region}} $\Omega^{\a}$ with radius $R^\a$, the {{\it blending region}} $\Omega^\b$ with width $L^{\b}$, and the {{\it continuum region}} $\Omega^\c$. 
We define the set of core atoms $\L^{\a} := \L \cap \Omega^\a$ and the set of blended atoms $\L^\b := \L \cap \Omega^\b$. We introduce the smooth blending functions satisfying $\beta \in C^{2,1}(\R^d)$ with $\beta=0$ in $\Omega^\a$ and $\beta=1$ in $\Omega^\c$, which will be specified for each individual examples later. The energy functional of BGFC method that we consider in this work is given by
\begin{eqnarray}\label{eq:gfc}
\E^{\rm bgfc}_h(u) := \E^{\rm bqce}_h(u) - \big\<\delta \E^{\rm bqce}_{\rm hom}({\bf 0}), u\big\>,
\end{eqnarray}
where the blended energy-based quasi-continuum (BQCE) energy functional reads
\begin{align}\label{eqn:Eqc}
  \E^{\rm bqce}_h(u) := \sum_{\ell \in \L^{\rm a}\cup\L^{\rm b}}& (1-\beta(\ell)) \cdot \Big(V_\ell\big(Du(\ell) + Du_0(\ell)\big) - V_\ell\big(Du_0(\ell)\big)\Big) \nonumber \\
  &+ \int_{\Omega}Q_{h} \Big(\beta(x)\cdot \big( W(\nabla u + \nabla u_0)-W(\nabla u_0)\big) \Big)\dx
\end{align}
with $Q_{h}$ being the $\mathcal{P}_0$ midpoint interpolation operator~\cite{2014-bqce} and $W : \R^{3 \times 3} \to \R$ being the Cauchy-Born energy density~\cite{e2007cb, ortner13}:
\begin{displaymath}
  W(\mF) := \det (\mA^{-1}) \cdot V(\mF \mathcal{R}).
\end{displaymath}
The homogeneous site potential $V$ and total energy functional $\E^{\rm bqce}_{\rm hom}$ are both evaluated on the homogeneous lattice $\Lhom$, $\mathcal{R}$ is the interaction range.

The equilibrium for the BGFC method is obtained by solving
\begin{equation}\label{eq::pbgfc}
u^{\rm bgfc}_h \in \arg \min  \big\{ \E^{\rm bgfc}_h(u_h) ~\big|~ u_h \in \Us_h \big\}, 
\end{equation}
where the solution space $\Us_h$ is given by \eqref{eq:space-bgfc}. To simplify the implementation, we will utilize the $\mathcal{P}_1$ finite element method to discretize the Cauchy-Born model within the BGFC method throughout this study. Extending the method to higher-order finite elements would require significant additional effort and will be explored in future work. The {\it a priori} error estimates of the BGFC method, in terms of degrees of freedom, for different types of crystalline defects, will be provided in Theorem~\ref{thm:bgfc}, \ref{sec:app:A}.

\subsubsection{Error estimator and adaptive algorithm}
\label{sec:sub:sub:est}

After establishing the {\it a priori} error analysis, the next step is to consider the adaptive atomistic/continuum (a/c) coupling. The crucial aspect of the {\it a posteriori} analysis for a/c coupling methods is to demonstrate the following estimate, where the constant $C$ is independent of any model parameters. 
\[
\| \nabla u^{\rm a} - \nabla u^{\rm bgfc}_h\|_{L^2} \leq C \eta(u^{\rm bgfc}_h),
\]
where $\eta(\cdot)$ is the {\it a posteriori} error estimator. The residual-based error analysis, utilizing the stress tensor formulation, offers valuable insights for dynamically adjusting the position of the atomistic/continuum (a/c) interface and adapting the discretization of the continuum region. This can be achieved by assigning local contributions, such that $\eta(u_h^{\text{bgfc}}) = \sum_{T \in \mathcal{T}_h} \eta_T(u_h^{\text{bgfc}})$. Further details on this topic can be found in our recent works \cite{liao2018posteriori, wang2018posteriori, CMAME, wang2021posteriori}, which also explore extensions of adaptive quantum mechanics/molecular mechanics (QM/MM) coupling methods.

To ensure clarity and simplicity in our presentation, we adopt the heuristic gradient-based error estimator~\cite{fu2023adaptive} throughout this paper. For the BGFC method, the local error estimator is chosen as $\eta_T = \|\nabla u_h^{\text{bgfc}}\|_{L^2(T)}$, where $T \in \mathcal{T}_h$. It is important to note that a rigorous {\it a posteriori} error estimate for the BGFC method will be investigated in future research. The corresponding adaptive algorithm has already been discussed in~\cite{fu2023adaptive}, and for the sake of completeness, we include it here.

\begin{algorithm}[H]
\caption{Mesh refinement for BGFC method.}
\label{alg:refine} 
Prescribe $0<\tau_1, \tau_2 <1$.  
\begin{enumerate}

    \item Given a coupled mesh $\T_h$ and the approximate solution $u^{\rm bgfc}_h$, compute the local error estimator $\eta_T = \|\nabla u_h^{\rm bgfc}\|_{L^{2}(T)}$ on each element $T \in \T_h$.

	\item Choose a minimal subset $\mathcal{M}\subset \T_h$ such that
	\begin{eqnarray}\label{eq:do}
		\sum_{T\in\mathcal{M}}\eta_{T}\geq\tau_1\sum_{T\in\T_h}\eta_{T}.
	\end{eqnarray}

	\item We can find the interface elements within $p$ layers of lattice spacing,  $\mathcal{M}_p:=\{T\in\mathcal{M}\bigcap(\T^{\b} \cup \T^{\c}): \textrm{dist}(T, \L^{\a})\leq p\}$. Choose $P>1$, find the first $p\leq P$ such that
	\begin{equation*}
		\sum_{T\in \mathcal{M}_p}\eta_{T}\geq \tau_2\sum_{T\in\mathcal{M}}\eta_{T},
		\label{eq:interface1}
	\end{equation*}
    \item Expand the atomistic region $\L^\a$ and the blending region $\Lambda^\b$ outward by $[\frac{p}{2}]$ and $p-[\frac{p}{2}]$ layers respectively. Bisect all elements $T\in  \mathcal{M}\setminus \mathcal{M}_p$ to obtain new triangulation $\T_h^{\rm new}$. 
\end{enumerate}
\end{algorithm}

\subsection{Edge dislocation in BCC W}
\label{sec:sub:edge}

We first consider the example of a straight edge dislocation, which is a common type of linear defect in crystalline materials that significantly affects their mechanical properties. To construct the dislocation, we adopt the same setup as presented in \cite{fu2023adaptive} for the body-centered cubic (BCC) tungsten crystal, using a quasi-2D approach \cite{mazars2011long}. In this case, a 3D simulation is performed, where clamped boundary conditions are imposed on the x-y plane, while periodic boundary conditions are enforced in the z-direction. To fit the clamped boundary conditions, three layers of ghost atoms are employed, and to improve the performance of the geometry optimization, these ghost atom layers on the x-y plane are relaxed.

\subsubsection{Overall efficiency and robustness}

\begin{figure}[H]
	\centering 
	\includegraphics[height=6cm]{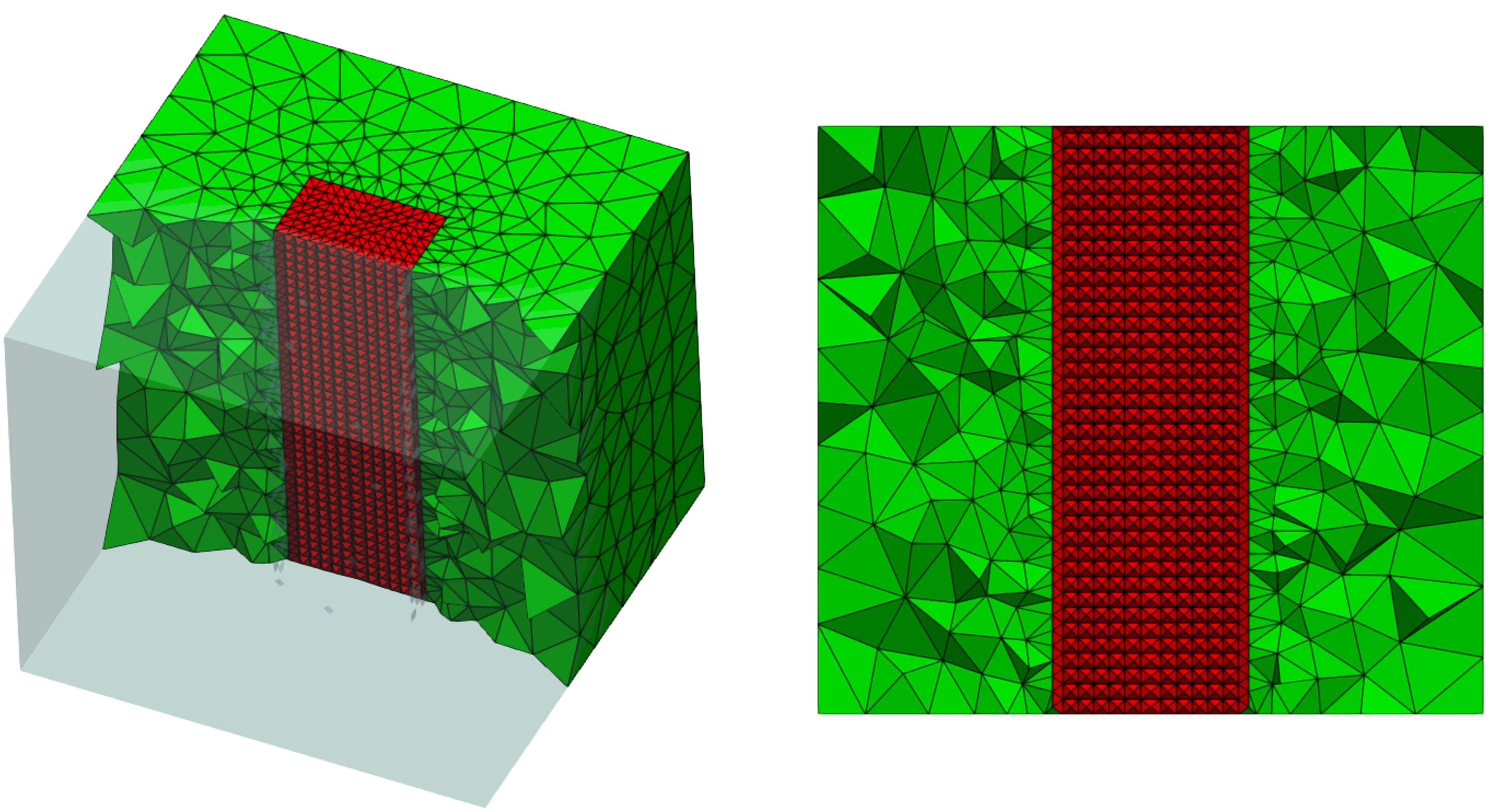}
	\caption{The illustration of the coupled mesh $\T_h$ for edge dislocation in BCC W constructed by {\tt MeshAC}.}
	\label{fig:edge_dislocation}
\end{figure}

The process of constructing the coupled mesh $\T_h$ for the (001)[100] edge dislocation in tungsten, as depicted in Figure~\ref{fig:edge_dislocation}, is facilitated by the mesh generation features integrated in {\tt MeshAC}. The atom positions are used as input to the algorithm, which then automatically generates the finite element mesh to accurately capture the behavior of the system. 

In Table~\ref{tab:disloc}, we present the computational times, measured in seconds, for solving the BGFC solution (cf. Equation~\ref{eq::pbgfc}), estimating the gradient-based error estimator, and refining the coupled mesh to achieve different degrees of freedom (DoF) during the adaptive computations for the edge dislocation in BCC W. It is evident that the cost associated with mesh generation and adaptation is minimal compared to the optimization process for the BGFC method on each coarse level. Specifically, the CPU time dedicated to mesh generation and refinement constitutes less than 5\% of the total CPU time.
\begin{table}[htbp]
\centering
\begin{tabular}{|c|c|c|c|c|}
\hline
Step & DoF & Solving time & Estimating time & Refining time \\ \hline
1  & 14286 & 742.22 & 221.76 & \textbf{26.43} \\ 
2  & 17826 & 989.25 & 365.29 & \textbf{35.41} \\ 
3  & 23572 & 1596.35 & 797.08 & \textbf{54.55} \\ 
4  & 28740 & 3981.53 & 1328.52 & \textbf{68.06} \\ 
\hline
\end{tabular}
\caption{CPU time(s) during adaptive process for edge dislocation in BCC W. Bold font indicates the efficiency of {\tt MeshAC} implementation.}
\label{tab:disloc}
\end{table}

\begin{figure}[htb]
	\centering 
	\subfigure[Geometry error]{
	\includegraphics[height=6cm]{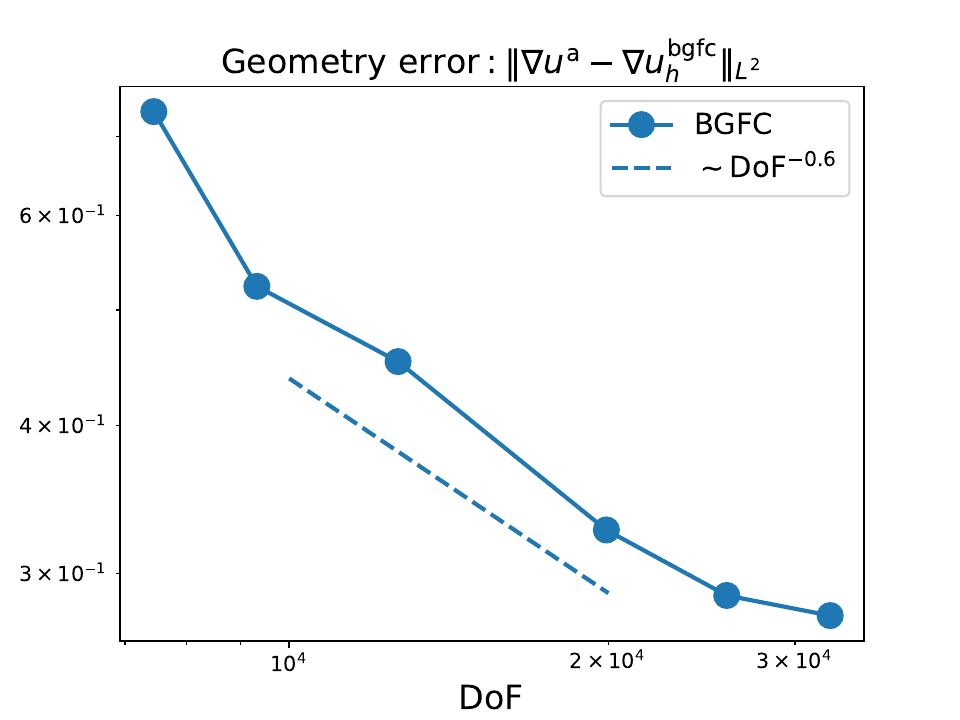}}~
	\subfigure[Energy error]{
	\includegraphics[height=6cm]{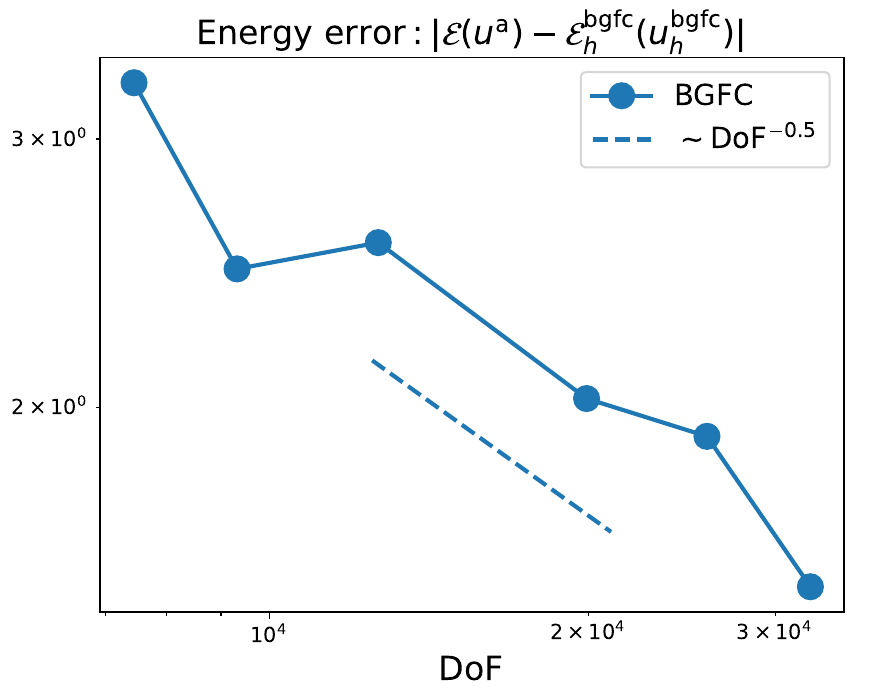}} 
 \caption{Errors vs. DoF for edge dislocation in BCC W.}
	\label{fig:conv_edge_dislocation}
\end{figure}

Figure \ref{fig:conv_edge_dislocation} showcases the convergence behavior of the geometry error $\|\nabla u^{\rm a} - \nabla u^{\rm bgfc}_h\|_{L^2}$ and the energy error $|\E(u^{\rm a}) - \E^{\rm bgfc}_h(u^{\rm bgfc}_h)|$ as a function of the number of degrees of freedom (DoF) during the adaptive computations of the BGFC solution for the (001)[100] edge dislocation in BCC tungsten. Both sub-figures exhibit nearly half-order convergence rates, which, while not achieving the optimality described in Proposition~\ref{thm:bgfc}, represent a significant finding as the convergence analysis for complex defects beyond point defects has not been explored extensively in the context of adaptive multiscale coupling, to the best of our knowledge. We attribute this sub-optimality to the use of a heuristic gradient-based error estimator, rather than the more reliable residual-based error estimator which is known to provide upper bounds on the approximation error and leads to optimal first order convergence rate in 2D ~\cite{wang2018posteriori}. In future work, we plan to investigate the development of a rigorous {\it a posteriori} error estimate for the BGFC method and incorporate this functionality into {\tt MeshAC}. Overall, these results demonstrate the accuracy of our package and offer valuable insights into the potential for precise multiscale simulations of complex crystalline defects.

\subsubsection{Adaptation process}

\begin{figure}[H]
	\centering 
	\subfigure[Iteration 1]{
	\includegraphics[scale=0.35]{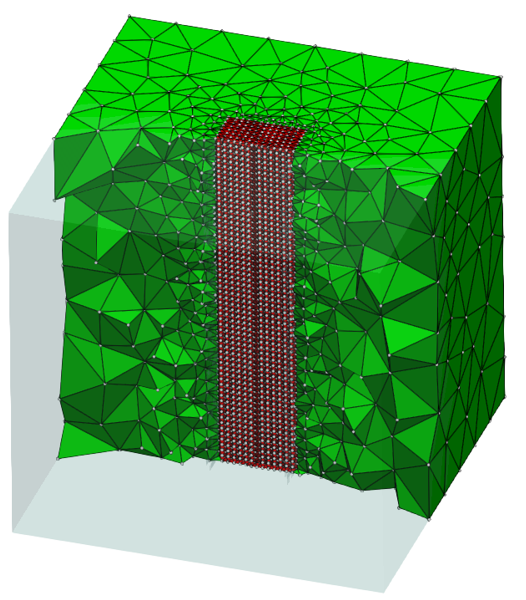}} \qquad
	\subfigure[Iteration 2]{
	\includegraphics[scale=0.35]{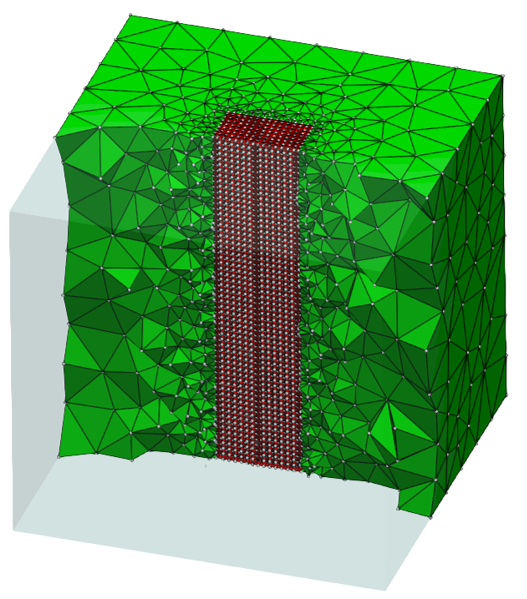}} \\
	\subfigure[Iteration 3]{
	\includegraphics[scale=0.35]{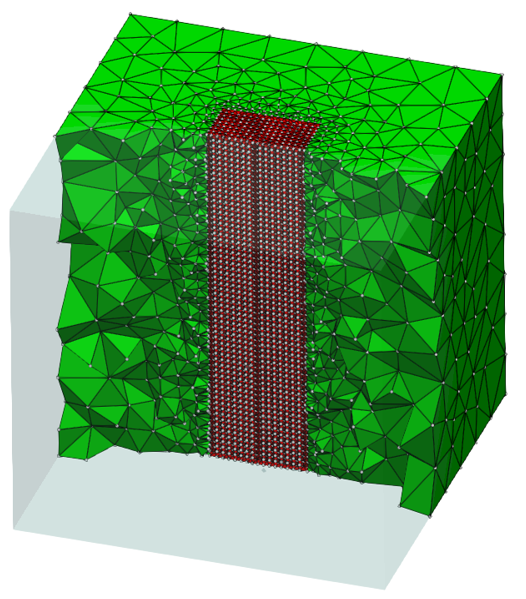}} \qquad
	\subfigure[Iteration 4]{
	\includegraphics[scale=0.35]{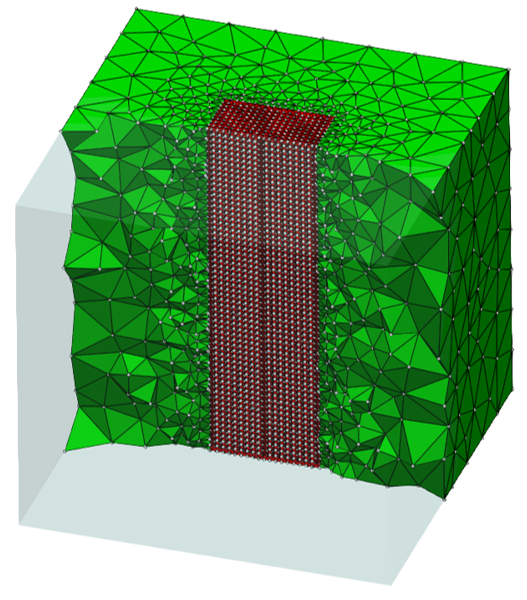}}
	\caption{Adaptive mesh refinement for edge dislocation in BCC W.}
	\label{fig:edge_dislocation_steps}
\end{figure}

\begin{figure}[H]
	\centering 
	\includegraphics[height=7cm]{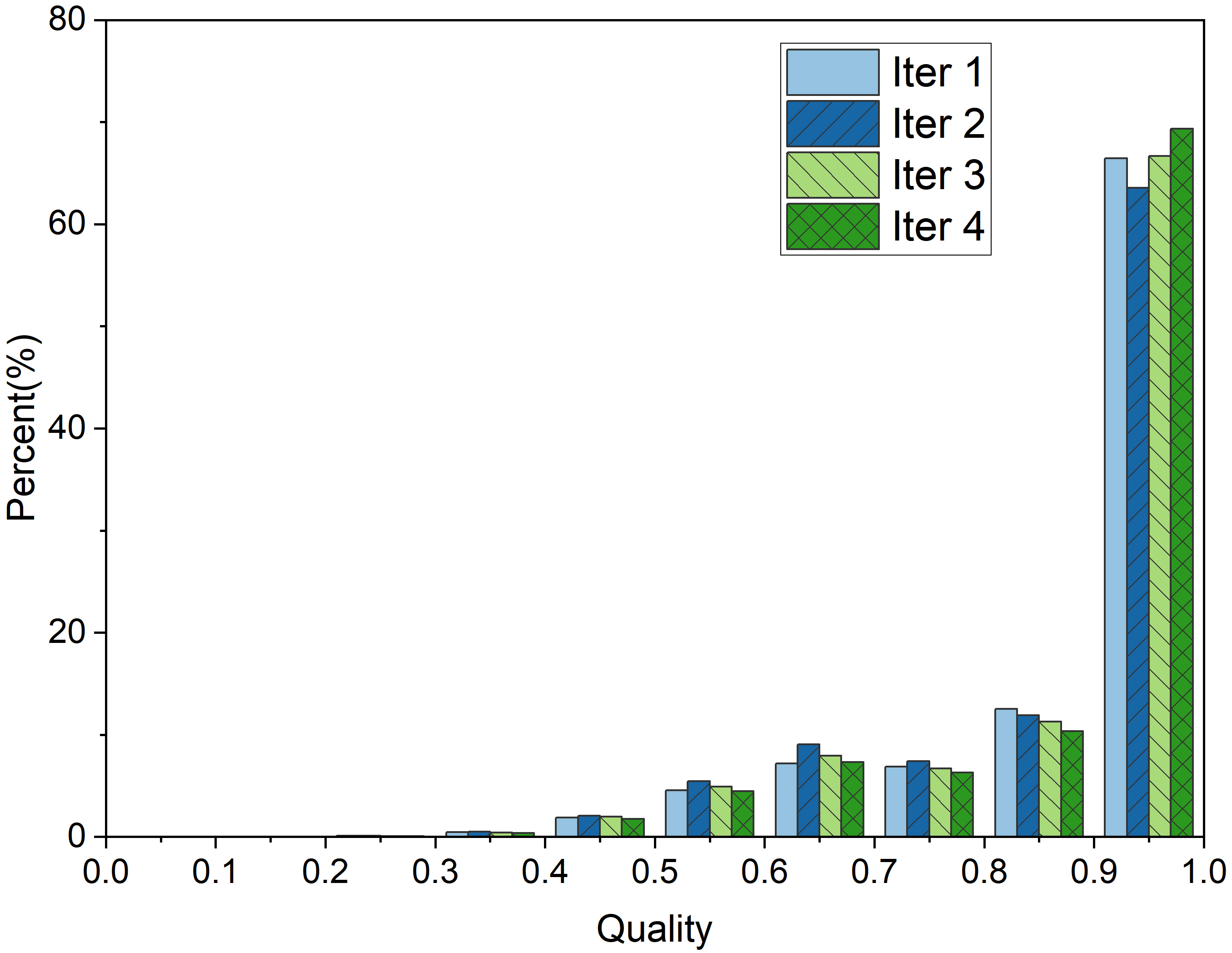}
	\caption{Mesh quality statics during mesh refinement for for edge dislocation in BCC W.}
	\label{fig:disloc_quality}
\end{figure}

The initial four iteration steps of the adaptive mesh refinement process for the edge dislocation in BCC W are illustrated in Figure~\ref{fig:edge_dislocation_steps}. To evaluate and analyze mesh quality during the refinement iterations, we utilize the quality metric defined in Equation \eqref{eq:element_quality}. The results, as depicted in Figure~\ref{fig:disloc_quality}, demonstrate consistent maintenance of mesh quality throughout the refinement process, effectively avoiding any detrimental effects that can arise from deteriorating mesh quality on the accuracy of computational results. The examined meshes exhibit a proportion of high-quality elements ($q(K)\in (0.9, 1.0]$) of $66.47\%$, $63.56\%$, $66.70\%$, and $69.36\%$, respectively. While the boundary conditions employed for the edge dislocation may have obscured any noticeable improvements in mesh quality during the refinement iterations, these findings still provide evidence of the effectiveness and robustness of {\tt MeshAC}. It is reasonable to anticipate that further improvements in mesh quality will become more evident with careful treatment of the boundary conditions.

Those results strongly support the efficiency and robustness of our implementation of {\tt MeshAC}, enabling large-scale multiscale simulations for crystalline defects. 

\subsection{Double voids in FCC Cu}
\label{sec:sub:diinclu}

\begin{figure}[H]
	\centering 
	\includegraphics[height=7cm]{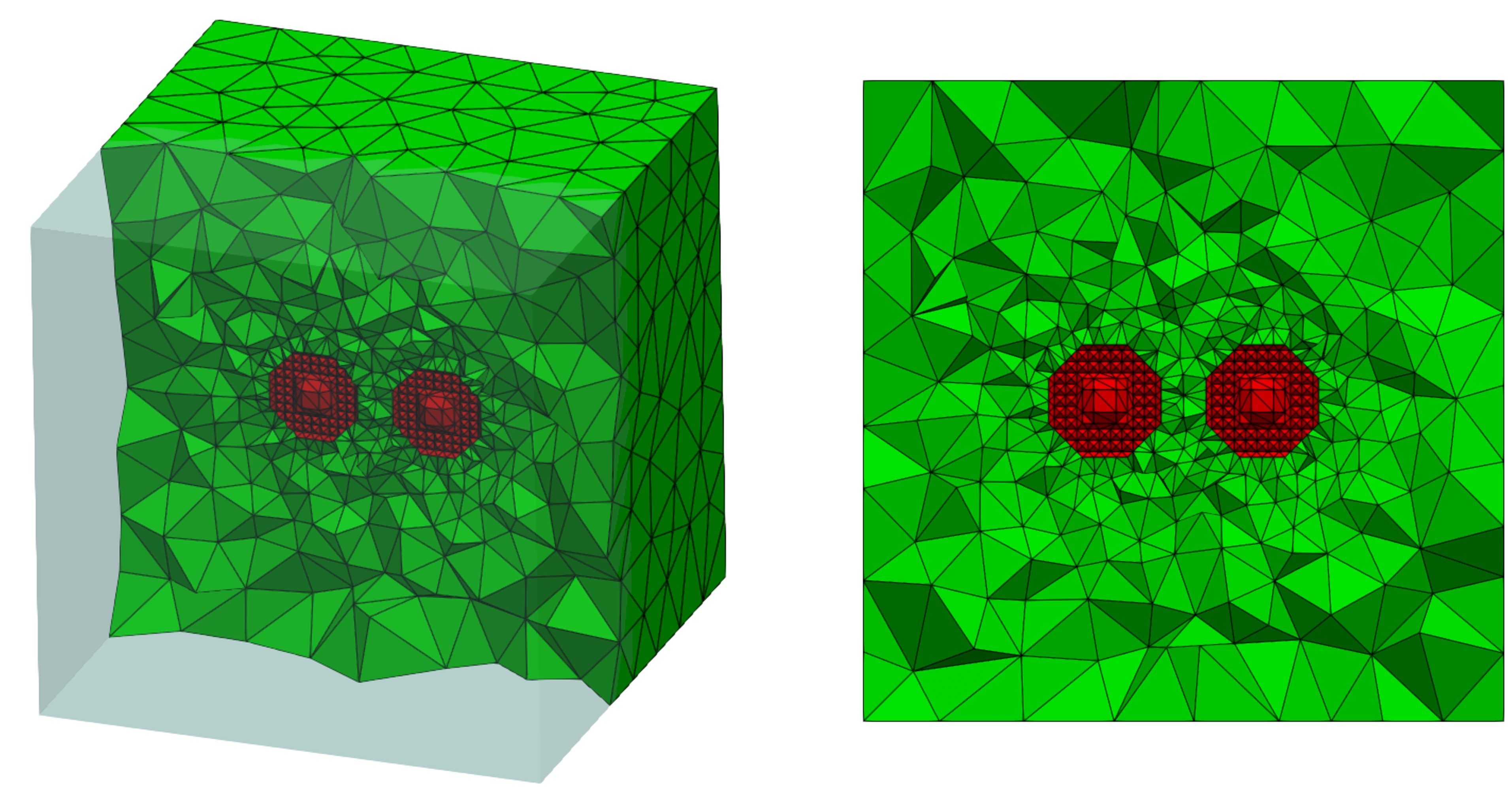}
	\caption{The illustration of the coupled mesh $\T_h$ for double voids in FCC Cu constructed by {\tt MeshAC}.}
	\label{fig:two_voids}
\end{figure}

The next numerical example we present is an extension of a previously studied (two-dimensional) case \cite{tembhekar2017automatic} to three dimensions. Specifically, we consider the double voids in face-centered cubic (FCC) Cu. Inclusions, depending on their size, shape, and distribution within the crystal lattice, can significantly influence the material's properties. Therefore, accurately modeling such defects using a/c coupling methods is crucial for enhancing our understanding of material failure and facilitating the design of novel materials. Figure~\ref{fig:two_voids} showcases the coupled mesh $\T_h$ for the double voids in FCC Cu, constructed using the capabilities of {\tt MeshAC}.

\begin{table}[H]
\centering
\begin{tabular}{|c|c|c|c|c|}
\hline
Step & DoF & Solving time & Estimating time & Refining time \\ \hline
1  & 2963 & 142.86 & 40.34 & \textbf{8.66} \\ 
2  & 4539 & 268.38 & 65.18 & \textbf{13.01} \\ 
3  & 6481 & 439.61 & 97.32 & \textbf{20.52} \\ 
4  & 8324 & 931.66 & 128.98 & \textbf{30.52} \\ 
\hline
\end{tabular}
\caption{CPU time(s) during adaptive computations for double voids in FCC Cu. Bold font indicates the efficiency of {\tt MeshAC} implementation.}
\label{tab:diinclusion}
\end{table}

Similar to the edge dislocation example presented in the previous section, Table~\ref{tab:diinclusion} showcases the CPU times in seconds during the adaptive computations for the double voids in FCC Cu. It is important to note that the computational cost associated with mesh generation and adaptation remains minimal compared to the optimization process for the BGFC method. Specifically, the proportion of CPU time dedicated to mesh generation and refinement is once again less than 5\% of the total CPU time. This highlights the efficiency of our mesh generator \cite{CPCmesh} in effectively handling various types of defects, further emphasizing its suitability for large-scale simulations.

\begin{figure}[H]
	\centering 
	\subfigure[Geometry error]{
	\includegraphics[height=6cm]{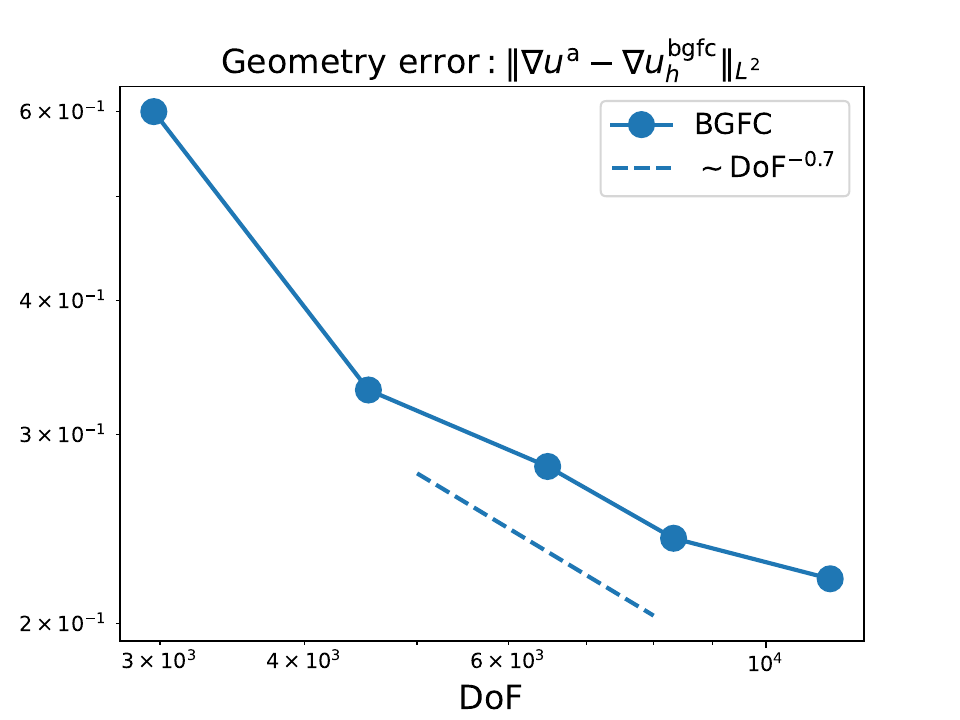}}~
	\subfigure[Energy error]{
	\includegraphics[height=6cm]{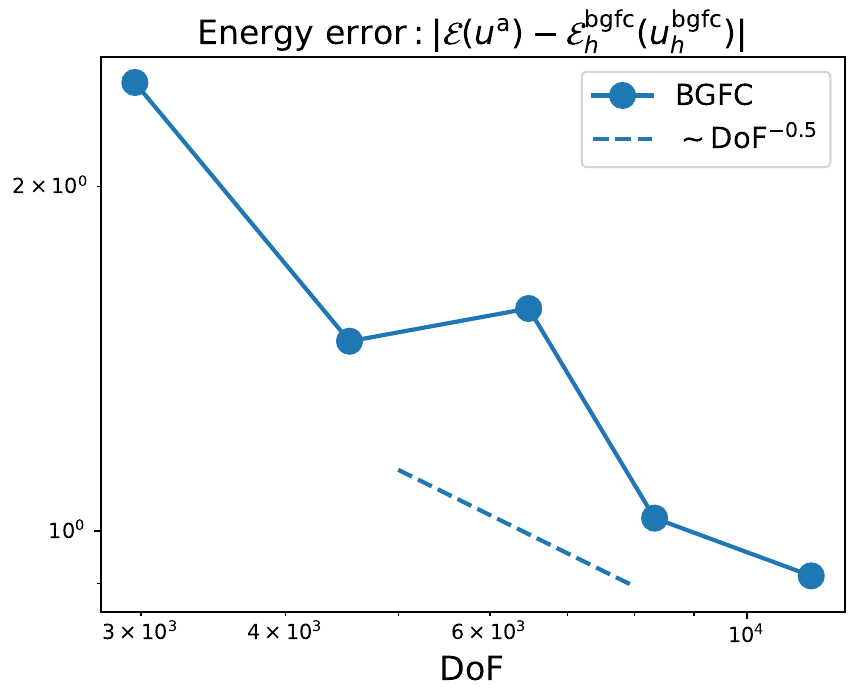}} 
 \caption{Errors vs. DoF for double voids in FCC Cu.}
	\label{fig:conv_diinclusion}
\end{figure} 

Figure~\ref{fig:conv_edge_dislocation} presents the convergence behavior of the geometry error $\|\nabla u^{\rm a} - \nabla u^{\rm bgfc}_h\|_{L^2}$ and the energy error $|\E(u^{\rm a}) - \E^{\rm bgfc}_h(u^{\rm bgfc}_h)|$ with respect to the number of degrees of freedom (DoF) in the adaptive computations of the BGFC solution for the double voids in Cu. The geometry error exhibits a decay rate of approximately $\textrm{DoF}^{-0.7}$, while the energy error converges at a nearly half-order rate. It is important to note that, currently, there is no theoretical result available for this specific type of defect. However, we anticipate that the convergence rates will improve once a residual-based error estimator is developed and incorporated into the BGFC method.

\begin{figure}[H]
	\centering 
	\subfigure[Iteration 1]{
	\includegraphics[scale=0.4]{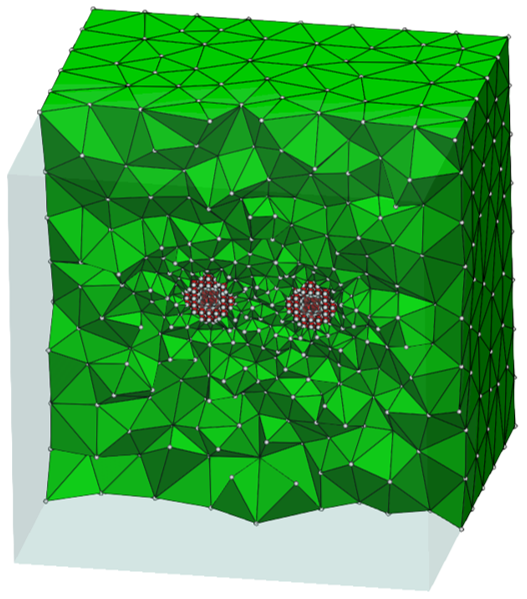}} \qquad
	\subfigure[Iteration 2]{
	\includegraphics[scale=0.4]{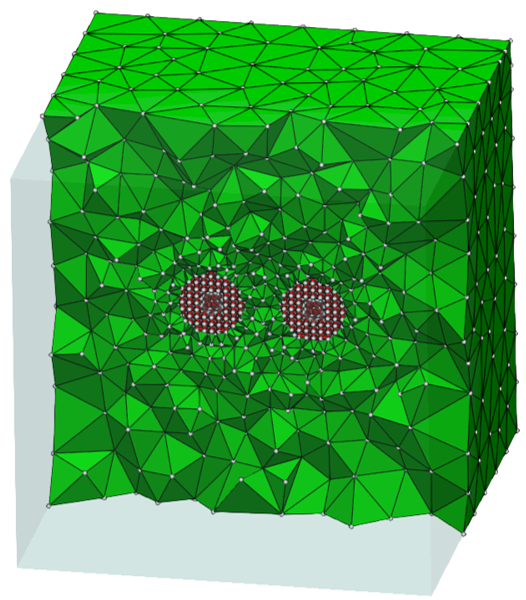}} \\
	\subfigure[Iteration 3]{
	\includegraphics[scale=0.4]{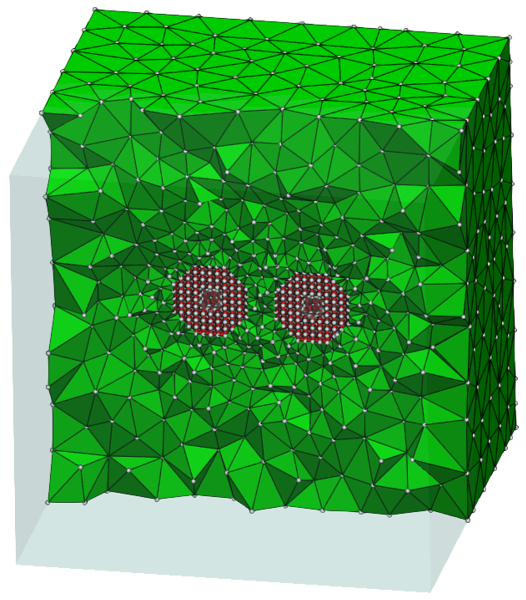}} \qquad
	\subfigure[Iteration 4]{
	\includegraphics[scale=0.4]{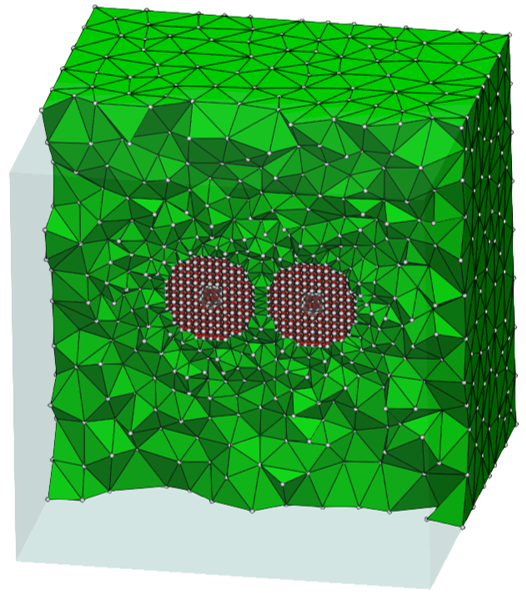}}
	\caption{Adaptive mesh refinement for the double voids in FCC Cu.}
	\label{fig:two_voids_steps}
\end{figure}

\begin{figure}[H]
	\centering 
	\includegraphics[height=7.5cm]{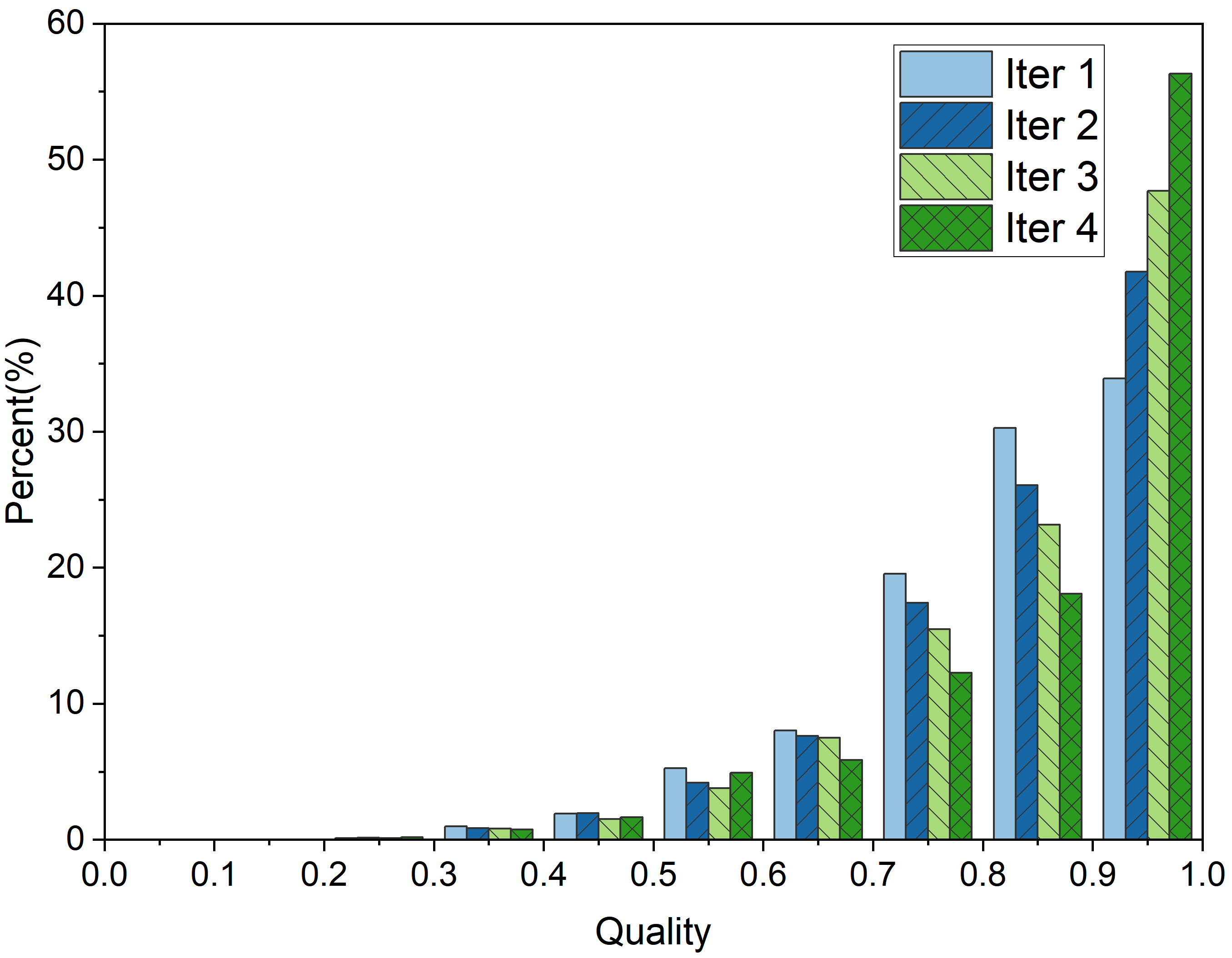}
	\caption{Mesh quality statics during mesh refinement for the double voids in FCC Cu.}
	\label{fig:diinclusion_quality}
\end{figure}

Figure \ref{fig:two_voids_steps} showcases the initial four iteration steps of the adaptive mesh refinement process for the double voids in FCC Cu. The quality of the mesh is evaluated using the metric defined in \eqref{eq:element_quality}, and the results are depicted in Figure~\ref{fig:diinclusion_quality}. These results clearly illustrate the improvement in mesh quality throughout the refinement process, leading to enhanced accuracy of the computational results. The examined meshes at different refinement levels for the double voids case exhibit a noticeable increase in the proportion of high-quality elements ($q(K)\in (0.9, 1.0]$), which constitute $33.92\%$, $41.76\%$, $47.71\%$, and $56.33\%$ of the meshes, respectively. These findings provide compelling evidence for the effectiveness and robustness of the implementation of {\tt MeshAC}.

\subsection{More extensions for mesh generation}
\label{sec:sub:ext}

This section aims to showcase the versatility of our package, {\tt MeshAC}, by demonstrating its capabilities in handling complex crystalline defects. We expand the functionality of a/c mesh generation by presenting additional examples that utilize tungsten (W) as the chosen metal element. Tungsten is a well-studied material in the field of materials science, which enhances the relevance and applicability of our demonstrations. However, it is important to note that our package is not limited to tungsten and can readily accommodate other elements without any technical complications. This flexibility allows researchers and practitioners to explore a wide range of materials and defects using {\tt MeshAC}.

\begin{figure}[H]
	\centering 
\includegraphics[height=7cm]{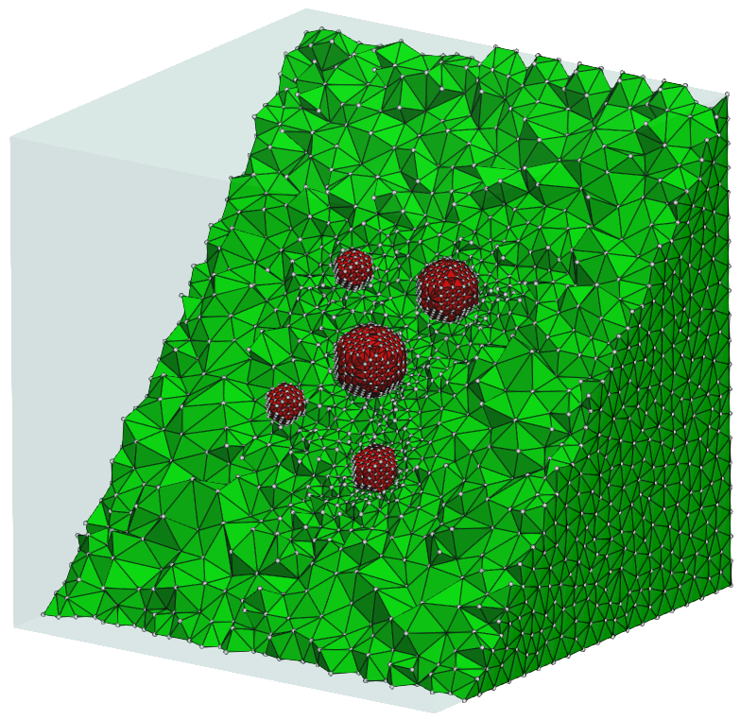}
	\caption{The illustration of a/c mesh for multiple holes in BCC W.}
	\label{fig:multi_holes}
\end{figure}

In the first example, we explore the capabilities of mesh generation for configurations involving multiple holes, building upon the double voids case discussed in Section~\ref{sec:sub:diinclu}. Figure~\ref{fig:multi_holes} illustrates this scenario, which presents significant challenges due to the increased geometric complexity. Constructing an atomistic-to-continuum mesh that accurately captures the multiple holes is a non-trivial task. However, our package {\tt MeshAC} effortlessly handles this complexity, enabling the generation of high-quality meshes.

\begin{figure}[H]
	\centering \includegraphics[height=6cm]{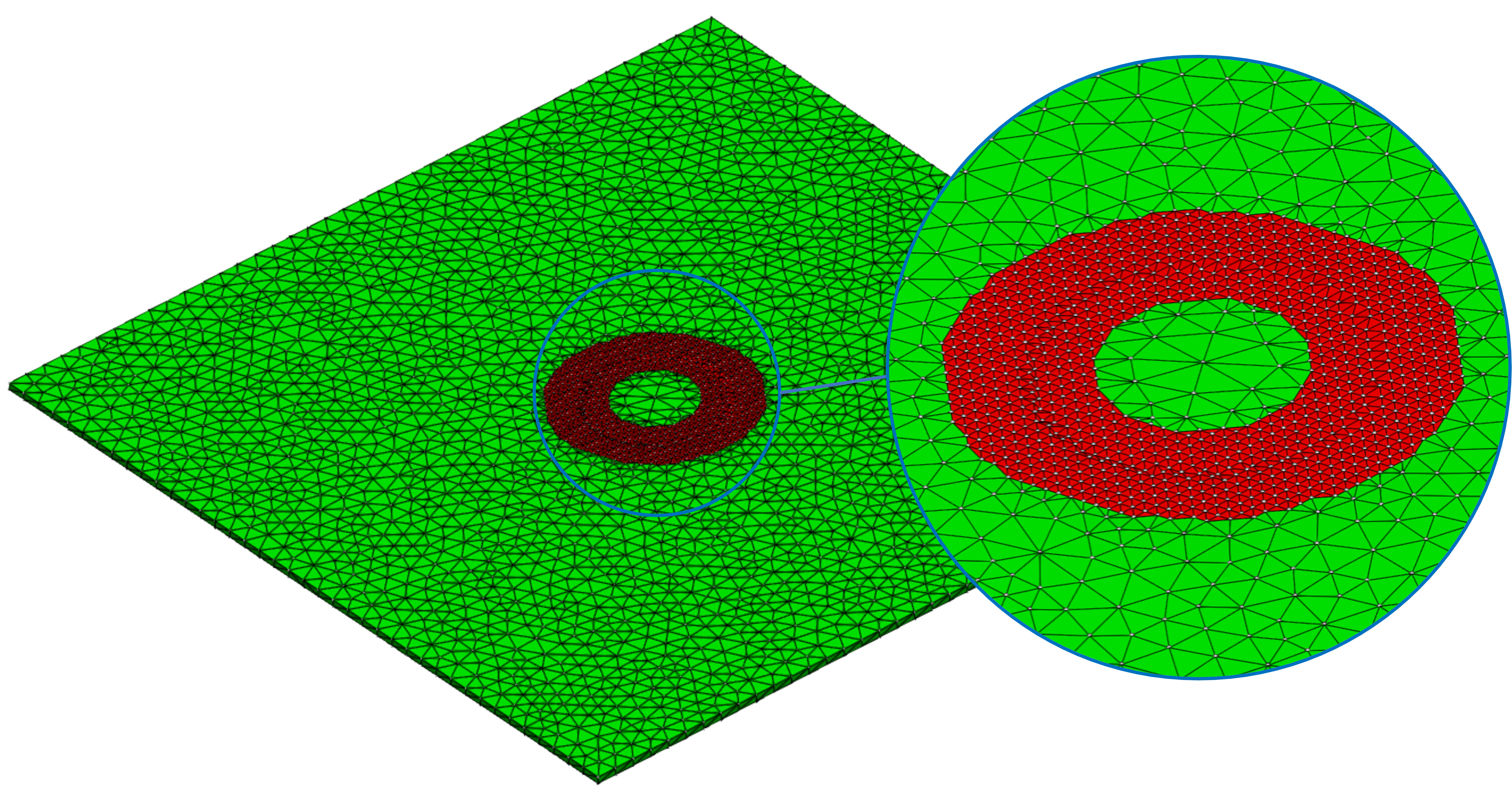}
	\caption{The illustration of a/c mesh for a dislocation loop in BCC W.}
	\label{fig:loop_max}
\end{figure}

The second example, shown in Figure~\ref{fig:loop_max}, focuses on a dislocation loop, where the atomistic configuration is generated using the {\tt Atomsk} tool \cite{hirel2015atomsk}. Constructing a mesh that accurately represents the intricate details of dislocation loops is a challenging endeavor. Nevertheless, our package {\tt MeshAC} excels in capturing these intricate loop structures, making the mesh generation process seamless.

\begin{figure}[H]
	\centering \includegraphics[height=6cm]{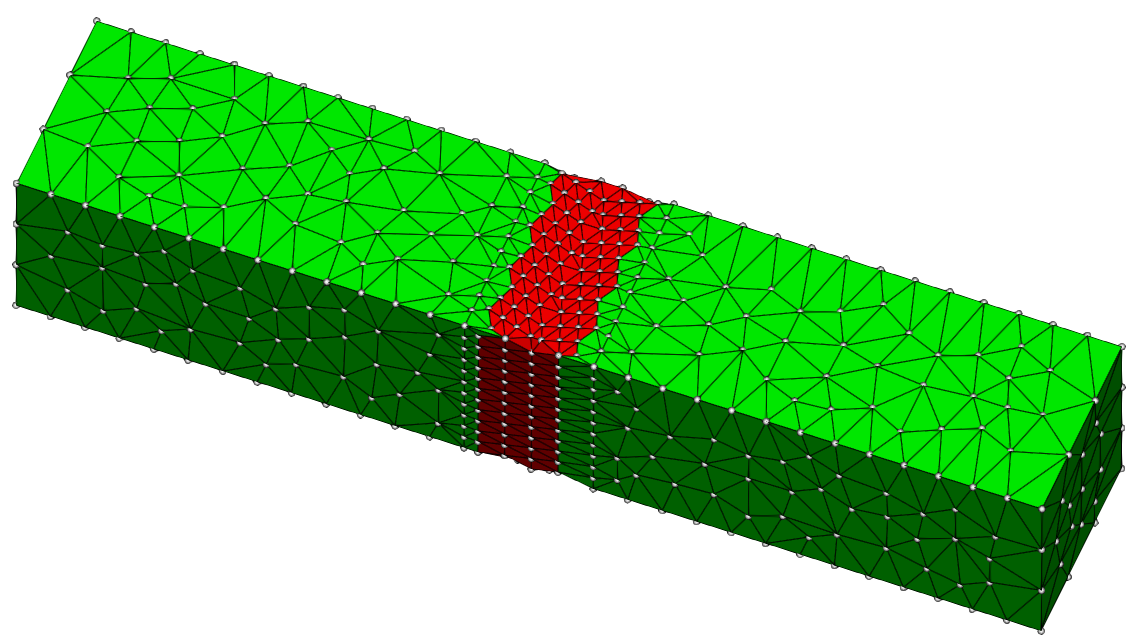}
	\caption{The illustration of a/c mesh for a $\Sigma_5$ grain boundary in BCC W.}
	\label{fig:gb}
\end{figure}

In the third example, we assess the performance of our package, {\tt MeshAC}, on a planar defect type known as the $\Sigma_5$ grain boundary, as depicted in Figure~\ref{fig:gb}. Constructing an atomistic-to-continuum mesh for grain boundaries has historically been a challenging task due to their intricate nature. However, with the assistance of our package {\tt MeshAC}, the construction of such meshes becomes more manageable, highlighting its effectiveness in overcoming the challenges associated with diverse crystalline defects.

\section{Conclusion}
\label{sec:con}

The software package {\tt MeshAC} has been specifically developed to address the challenges of implementing effective atomistic/continuum (a/c) coupling methods in three dimensions. It accomplishes this by integrating advanced mesh generation and adaptation features. The package is designed to respect the crystalline structure and the configuration of defects. Also, it incorporates localized modification and reconstruction capabilities to facilitate the operation of the a/c interface, thereby enhancing both implementation efficiency and the quality of the coupled mesh. During simulation, {\tt MeshAC} extends the atomistic region and adapts the coupled mesh to achieve optimal accuracy and efficiency using a heuristic gradient-based {\it a posteriori} error estimator. The effectiveness of {\tt MeshAC} has been demonstrated through the application of the BGFC method for practical material defects, which  The results demonstrate that the accuracy and efficiency of {\tt MeshAC} make it a valuable tool for the computational mechanics community. While our presentation has focused on mesh refinement and static a/c coupling methods, we acknowledge that extending the methodology to mesh coarse-graining and moving mesh requires further technical development. Nonetheless, we believe that there is no fundamental limitation to this extension and we plan to address these features and add more functionals such as rigorous residual-based estimators and more robust adaptive algorithms to {\tt MeshAC} in our future work.

\appendix

\section{Preliminaries}
\label{sec:app:A}

In this section, we introduce the fundamental concepts necessary for understanding Section~\ref{sec:sub:bgfc}, following the results of previous studies \cite{colz2016, fang2020blended}. To maintain conciseness, we primarily focus on single-species Bravais lattices. However, it should be emphasized that the techniques presented in this work can be extended to multilattice crystals \cite{olson2019theoretical} with minor adjustments to the notations.

\def\Rcore{R_{\rm DEF}}
\def\UsH{{\mathscr{U}}^{1,2}}
\def\Adm{{\rm Adm}}

We denote by $\Us := \{v: \L\to \mathbb{R}^3 \}$ the set of vector-valued lattice functions. Recall that the deformed configuration of $\L$ is a map $y\in \Us$ which can be decomposed as
\begin{eqnarray}\label{y-u}
y(\ell) = \ell + u_0(\ell) + u(\ell) \qquad\forall~\ell\in\Lambda,
\end{eqnarray}
where $u_0: \L\rightarrow\R^3$ is a {\it far-field predictor} enforcing the presence of the defect of interest and $u: \L\rightarrow\R^3$ is a {\it corrector}.
For point defects, we simply take $u_0=0$. For straight dislocation lines as well as anisotropic cracks, $u_0$ can be derived by solving a continuum linearized elasticity (CLE) equation and we refer to~\cite{2013-defects, 2019-crackbif, 2021-defectexpansion, gen2dislocation} for more details.  

For each atom $\ell\in \L$ and $v \in \Us$, we define the finite difference stencil $Dv(\ell):= \{D_\rho v(\ell)\}_{\rho \in \Rg_\ell} :=\{v(\ell+\rho)-v(\ell)\}_{\rho \in \Rg_\ell}$,
where $\Rg_\ell := \{\ell'-\ell~|~\ell'\in \Nhd_\ell\}$ is the interaction range with interaction neighborhood $\Nhd_{\ell} := \{ \ell' \in \L~|~0<|\ell'-\ell| \leq \rcut \}$ with some cut-off radius $\rcut>0$.

For $u\in\Us$, its nodal interpolant with respect to the background mesh is denoted as $\bar{u}$~\cite{2014-bqce, colz2016, gen1}. We then introduce the discrete homogeneous Sobolev spaces
\begin{eqnarray}\label{eq:spce-a}
	\Use :=\{u\in \Us ~|~\nabla \bar{u}\in L^2\} \quad \textrm{with semi-norm}~|\cdot|_{\Use}:=\|\nabla \bar{u}\|_{L^2}.~
\end{eqnarray}

The site potential is a collection of mappings $V_{\ell}:(\R^3)^{\Rg_\ell}\rightarrow\R$, which represents the energy distributed to each atomic site in $\L$. We refer to \cite[Section 2]{2013-defects} for the detailed discussions on the assumptions of general site potentials. In this work, we will use the well-known EAM (Embedded Atom Method) model \cite{Daw1984a} throughout the numerical experiments (cf.~Section~\ref{sec:numer}).

We can formally define the energy-difference functional of the atomistic model
\begin{align*}
\E(u) =& \sum_{\ell\in\Lambda}\Big(V_{\ell}\big(Du_0(\ell)+Du(\ell)\big)-V_{\ell}\big(Du_0(\ell)\big)\Big).
\quad
\end{align*}
The equilibrium is obtained by solving the following minimization problem
\begin{eqnarray}\label{eq:variational-A-problem}
u^{\rm a} \in \arg \min \big\{\E(u) ~\big|~ u \in \Use \big\}.
\end{eqnarray}
where ``$\arg\min$'' is understood as the set of local minima.




Before introducing the BGFC method, we first present the basic notions of Cauchy-Born approximation. To reduce {\rm DoF} and meanwhile preserve considerable accuracy, we define a continuum approximation by the Cauchy-Born rule which is a typical choice in the multiscale context~\cite{e2007cb, ortner13}. The Cauchy-Born energy density functional $W : \R^{3 \times 3} \to \R$ reads
\begin{displaymath}
  W(\mF) := \det (\mA^{-1}) \cdot V(\mF \mathcal{R}),
\end{displaymath}
where $V$ is the homogeneous site energy potential on $\Lhom$.

For BGFC method, the space of coarse-grained displacements is defined as  
\begin{align}\label{eq:space-bgfc}
  \Us_h := \big\{ u_h \in C^0(\R^3;\R^3) ~\big|~
  \text{ $u_h$ is p.w. affine w.r.t. $\T_h$,} \text{ $u_h = 0$ in $\R^3 \setminus \Omega$ } \big\}.
\end{align}
The coarse-grid problem for the BGFC method reads
\begin{equation}\label{eq:variational-BGFC-problem}
u^{\rm bgfc}_h \in \arg \min  \big\{ \E^{\rm bgfc}_h(u_h) ~\big|~ u_h \in \Us_h \big\}. 
\end{equation}

The subsequent lemma provides the {\it a priori} error estimates of the BGFC method, in terms of the degrees of freedom ({\rm DoF}), for a point defect in three dimensions, a straight dislocation, and a crack. These estimates are based on \cite[Section 4.2.1]{colz2016} and \cite[Theorem 2.1]{fang20}, respectively. This offers the foundation of the {\it a posteriori} error estimates of the corresponding a/c coupling method.

\begin{lemma}\label{thm:bgfc}
Suppose that the blending function $\beta$ and the triangulation $\T_h$ satisfy \cite[Assumption 1]{fang20}, and $\mathcal{P}_1$ finite element method is applied in the continuum region, we have
\begin{align*}
    \|\nabla u^{\rm a} - \nabla u^{\rm bgfc}_h\|_{L^2} &\lesssim {\rm DoF}^{-5/6}, \qquad\qquad\qquad~~~~ \textrm{for 3D point defect}, \\
    \|\nabla u^{\rm a} - \nabla u^{\rm bgfc}_h\|_{L^2} &\lesssim {\rm DoF}^{-1} \cdot \log^{1/2}({\rm DoF}), \quad~~~~ \textrm{for dislocation}, \\
    \|\nabla u^{\rm a} - \nabla u^{\rm bgfc}_h\|_{L^2} &\lesssim {\rm DoF}^{-1/4} \cdot \log^{1/4}({\rm DoF}), \quad~~ \textrm{for crack},
\end{align*}
where $u^{\rm a}$ and $u^{\rm bgfc}_h$ are the solutions of \eqref{eq:variational-A-problem} and \eqref{eq:variational-BGFC-problem} respectively.
\end{lemma}

Next, we present the {\it a posteriori} error estimate. Let $I_{\a}$ be a piecewise interpolant defined on lattice~\cite{2013-defects}, the residual $\mR(I_{\a}u_h)$ as an operator on $\Use$ is defined by 
\begin{equation}\label{eq:def_res}
\mR(I_{\rm a} u_{h})[v] := \< \del\Ea(\Ia u_{h}), v\>, \quad \forall v\in \Use.
\end{equation}

The following lemma characterizes the dual norm of the residual and has been presented in our previous works~\cite{ac23_1, ac23_2}. We include it here for the sake of completeness.

\begin{lemma}\label{lemma:res-F}
	Let $u^{\rm a}$ be a strongly stable solution of the atomistic model \eqref{eq:variational-A-problem}, then there exists a solution $u^{\rm bgfc}_{h}$ of \eqref{eq:variational-BGFC-problem}, and constants $c, C$ independent of the approximation parameters such that 
	\begin{eqnarray}\label{res-bound}
	c\|I_{\rm a} u^{\rm bgfc}_{h} - u^{\a}\|_{\UsH} \leq \| \mR(I_{\rm a} u^{\rm bgfc}_{h}) \|_{(\UsH)^*} \leq C\|I_{\rm a} u^{\rm bgfc}_{h} - u^{\a}\|_{\UsH},
	\end{eqnarray}
 where the residual $\textsf{R}(I_{\rm a} u_{h})$ is defined by \eqref{eq:def_res}
\end{lemma}

Based on this lemma, the {\it ideal a posteriori} error estimator reads
\[
\eta^{\rm ideal}(u^{\rm bgfc}_{h}) := \| \mR(I_{\rm a} u^{\rm bgfc}_{h}) \|_{(\UsH)^*},
\]
which provides both upper and lower bounds for the approximation error. However, computing this estimator in practice is not feasible, which poses a significant limitation in the context of {\it a posteriori} error control and adaptive algorithm design. To overcome this challenge, we have primarily pursued two approaches for deriving practical error estimators. One approach involves the utilization of a force-based error estimator, which provides an upper bound for $\eta^{\rm ideal}$ \cite{ac23_1}. The second approach employs a stress-based error estimator, which offers both upper and lower bounds \cite{ac23_2}. In the specific context of this paper, we have opted to utilize the gradient-based error estimator due to its simplicity and computational efficiency. Incorporating additional error estimators such as force-based and stress-based error estimators will enhance the flexibility and applicability of {\tt MeshAC} for a wider range of problems. This is an avenue that we intend to explore in our future work, to further improve the capabilities and performance of the package.



\bibliographystyle{elsarticle-num}
\bibliography{mesh}







\end{document}

%% file: notation.tex
\renewcommand{\cases}[1]{\left\{ \begin{array}{rl} #1 \end{array} \right.}
\newcommand{\smfrac}[2]{{\textstyle \frac{#1}{#2}}}
\newcommand{\myvec}[1]{\left[ \begin{vector} #1 \end{vector} \right]}
\newcommand{\mymat}[1]{\left[ \begin{matrix} #1 \end{matrix} \right]}

\def\Xint#1{\mathchoice
{\XXint\displaystyle\textstyle{#1}}%
{\XXint\textstyle\scriptstyle{#1}}%
{\XXint\scriptstyle\scriptscriptstyle{#1}}%
{\XXint\scriptscriptstyle\scriptscriptstyle{#1}}%
\!\int}
\def\XXint#1#2#3{{\setbox0=\hbox{$#1{#2#3}{\int}$ }
\vcenter{\hbox{$#2#3$ }}\kern-.6\wd0}}
\def\mint{\Xint-}

\def\b{\big}
\def\B{\Big}
\def\bg{\bigg}
\def\Bg{\Bigg}


\def\diam{{\textrm{diam}}}
\def\conv{{\textrm{conv}}}
\def\t{\top} 
\def\sign{{\textrm{sgn}}}
\def\id{{\textrm{id}}}
\def\supp{{\textrm{supp}}}
\def\diam{{\textrm{diam}}}

\def\R{\mathbb{R}}
\def\N{\mathbb{N}}
\def\Z{\mathbb{Z}}
\def\C{\mathbb{C}}
\def\bbV{\mathbb{V}}

\def\WW{W}
\def\CC{C}
\def\HH{H}
\def\LL{L}
\def\DD{{D}'}
\def\Ys{\mathscr{Y}}

\def\WWh{\dot{W}}
\def\Ycb{Y}
\def\WWhz{\dot{W}_0}
\def\Ycbz{\Ycb_0}

\def\dx{\,{\textrm{d}}x}
\def\dy{\,{\textrm{d}}y}
\def\dz{\,{\textrm{d}}z}
\def\dr{\,{\textrm{d}}r}
\def\dt{\,{\textrm{d}}t}
\def\ds{\,{\textrm{d}}s}
\def\dd{\textrm{d}}
\def\pp{\partial}
\def\dV{\,\textrm{dV}}
\def\dA{\,{\textrm{dA}}}
\def\db{\,{\textrm{db}}}
\def\dlam{\,{\textrm{d}}\lambda}

\def\<{\langle}
\def\>{\rangle}

\def\ol{\overline}
\def\ul{\underline}
\def\ot{\widetilde}
\newcommand{\ut}[1]{\underset{\widetilde{\hspace{2.5mm}}}{#1}}

\def\mA{{\textsf{A}}}
\def\mB{\textsf{B}}
\def\mC{\textsf{C}}
\def\mF{\textsf{F}}
\def\mG{\textsf{G}}
\def\mH{\textsf{H}}
\def\mI{\textsf{I}}
\def\mJ{\textsf{J}}
\def\mP{\textsf{P}}
\def\mQ{\textsf{Q}}
\def\mR{\textsf{R}}
\def\mM{\textsf{M}}
\def\mS{\textsf{S}}
\def\mO{\textsf{0}}
\def\mL{\textsf{L}}

\def\sym{\textsf{sym}}
\def\tr{\textsf{tr}}
\def\el{\textsf{el}}

\def\bfa{\textbm{a}}
\def\bfg{\textbm{g}}
\def\bfrho{\mathbm{\rho}}
\def\bfv{\textbm{v}}
\def\bfh{\textbm{h}}
\def\bfO{\textbm{0}}

\def\bbA{\mathbb{A}}
\def\bbB{\mathbb{B}}
\def\bbC{\mathbb{C}}
\def\bbI{\mathbb{I}}

\def\Hs{\mathcal{H}}


\newcommand{\transpose}{{\!\top}}
\newcommand{\Da}[1]{D_{\!#1}}
\newcommand{\Dc}[1]{\D_{#1}}
\def\D{\nabla}
\def\del{\delta}
\def\ddel{\delta^2}
\def\dddel{\delta^3}

\def\loc{\textrm{loc}}

\def\qc{\textrm{qc}}
\def\h{\textrm{h}}

\def\eps{\varepsilon}
\def\tot{\textrm{tot}}
\def\cb{\textrm{cb}}
\def\a{\textrm{a}}
\def\c{\textrm{c}}
\def\b{\textrm{b}}
\def\ac{\textrm{ac}}
\def\i{\textrm{i}}
\def\nn{\textrm{nn}}
\def\refl{\textrm{rfl}}
\def\qnl{\textrm{qnl}}
\def\stab{\textrm{stab}}
\def\conv{\textrm{conv}}
\def\supp{\textrm{supp}}

\def\L{\Lambda}
\def\Is{\mathcal{I}}
\def\oIs{\ol{\Is}}
\def\As{\mathcal{A}}
\def\Cs{\mathcal{C}}
\def\Fs{\mathcal{F}}
\def\Ks{\mathcal{K}}
\def\Us{\mathscr{U}}
\def\Usz{\Us_0}
\def\Ush{\dot{\Us}^{1,2}}
\def\Ushd{\dot{\Us}^{-1,2}}
\def\Usp{{\Us}^{1,p}}
\def\Usc{\Us^c}

\def\Bs{\mathcal{B}}
\def\Ls{\mathcal{L}}
\def\bbL{\mathbb{L}}

\def\yF{y_\mF}
\def\uF{u_\mF}

\def\E{\mathscr{E}}
\def\Ea{\E^\a}
\def\Eb{\E^\textrm{b}}
\def\Ha{H^\a}
\def\Ei{\E^\i}
\def\Ec{\E^\c}
\def\Eh{\E^\h}
\def\F{\mathscr{F}}
\def\Hc{H^\c}
\def\Eqnl{\E^\qnl}
\def\Hqnl{H^\qnl}
\def\Erefl{\E^\refl}
\def\Hrefl{H^\refl}
\def\Eac{\E^\ac}
\def\Hac{H^\ac}
\def\Estab{\E^\stab}
\def\Hstab{H^\stab}

\def\dW{W'}
\def\ddW{W''}

\def\RO{\mathcal{R}}

\def\Es{\Phi}

\def\Eatot{\E^\a_\tot}
\def\Eatot{\E^\c_\tot}

\def\Om{{\R^d}}
\def\Vi{V^\i}
\def\Vc{V^\c}
\def\Vs{\mathscr{V}}

\def\tily{\tilde y}
\def\tilz{\tilde z}
\def\tilu{\tilde u}
\def\tilv{\tilde v}
\def\tile{\tilde e}
\def\tilw{\tilde w}
\def\tilf{\tilde f}

\def\bary{y}
\def\barz{z}
\def\barv{v}
\def\barw{w}
\def\baru{u}
\def\bare{e}
\def\barf{f}


\def\ve{\varepsilon}
\def\L{\Lambda}
\def\La{\L^{\a}}
\def\Li{\L^{\i}}
\def\Lc{\L^{\c}}
\def\yd{y_0}
\def\Nhd{\mathcal{N}}
\def\rcut{r_{\textrm{cut}}}
\def\Rg{\mathscr{R}}
\def\Rgp{\Rg^{+}}
\def\vsig{\varsigma}
\def\T{\mathcal{T}}
\def\Tp{T^+}
\def\Tm{T^-}
\def\np{\nu^+}
\def\nm{\nu^-}
\def\T{\mathcal{T}}
\def\Th{\mathcal{T}_h}
\def\Te{\mathscr{T}_\varepsilon}
\def\Fc{\mathscr{F}_\c}
\def\Fi{\mathscr{F}_\i}
\def\Fh{\mathscr{F}_h}
\def\UsT{\Us_h}
\def\Ih{I_h}
\def\Ia{I_\a}
\def\Ie{I_{\varepsilon}}
\def\Tmu{\mathscr{T}_\mu}
\def\vor\textrm{vor}
\def\s{\sigma}
\def\sa{\sigma^\a}
\def\sc{\sigma^\c}
\def\Sa{\Sigma^\a}
\def\Sc{\Sigma^\c}
\def\Oma{\Omega^\a}
\def\PO{\textrm{P}_0}

\def\Rdef{R^{\textrm{def}}}
\def\Rg{\mathcal{R}}
\def\Rgnn{\mathcal{N}}
\def\rcut{r_{\textrm{cut}}}
\def\Lhom{\L^{\textrm{hom}}}
\def\Ddef{D^{\textrm{def}}}
\def\Ldef{\L^{\textrm{def}}}

\def\Rdef{R^{\textrm{def}}}
\def\Rg{\mathcal{R}}
\def\Rgnn{\mathcal{N}}
\def\rcut{r_{\textrm{cut}}}
\def\Lhom{\L^{\textrm{hom}}}
\def\Ddef{D^{\textrm{def}}}
\def\Ldef{\L^{\textrm{def}}}

\def\Nh{\mathcal{N}_h}
\def\Ush{\Us_h}
\def\Ra{R^\a}
\def\Rb{R^{\textrm{b}}}
\def\Rc{R^\c}
\def\Eb{\E^{\textrm{b}}}
\def\dof{{\textrm{DOF}}}
\def\Omh{\Omega_h}
\def\Thr{{\T_{h,R}}}
\def\vor{\textrm{vor}}
\def\Uhr{\Us_{h,R}}

\def\Ta{\T_\a}
\def\Th{\T_{\textrm{h}}}
\def\sh{\sigma^{\textrm{h}}}